\documentclass[12pt]{article}
\usepackage{stmaryrd}
\usepackage{amssymb}
\usepackage{graphics}
\usepackage{epsfig}
\usepackage{a4wide}
\usepackage{cite}
\usepackage{color}

\begin{document}

\title{Ion pseudoheating by low-frequency Alfv\'{e}n waves Revisited}
\author{Chuanfei Dong$^\mathrm{a}$\thanks{%
dcfy@umich.edu} and Nagendra Singh$^\mathrm{b}$\thanks{%
singhn@uah.edu} \\
{$^\mathrm{a}${\small Department of Atmospheric, Oceanic and Space Sciences,
University of Michigan,}} \\
{\small Ann Arbor, MI 48109, U.S.A.}\\
{$^\mathrm{b}${\small Department of Electrical and Computer Engineering, The
University of Alabama, }} \\
{\small Huntsville, AL 35899, U.S.A.}}

\date{}
\maketitle

\begin{abstract}

Pseudoheating of ions in the presence of Alfv\'{e}n waves is studied. 
We show that this process can be explained by $E\times B$ drift. The analytic solution 
obtained in this paper are quantitatively in accordance with previous results.
Our simulation results show that the Maxwellian distribution is broadened during
the pseudoheating; however, the shape of the broadening distribution function
depends on the number of wave modes (i.e., a wave spectrum or a monochromatic
dispersionless wave) and the initial thermal speed of ions ($v_{p}$). It is of
particular interests to find that the Maxwellian shape is more likely
to maintain during the pseudoheating under a wave spectrum compared with a 
monochromatic wave. It significantly improves our understanding of heating processes 
in interplanetary space where Alfv\'{e}nic turbulences exist pervasively.
Compared with a monochromatic Alfv\'{e}n wave, $E\times B$ drift
produces more energetic particles in a broad spectrum of Alfv\'{e}n
waves, especially when the Alfv\'{e}nic turbulence with phase coherent wave modes
is given. Such particles may escape from the region of interaction with the Alfv\'{e}n 
waves and can contribute to fast particle population in astrophysical
and space plasmas.
\end{abstract}

\vskip 15mm


\renewcommand{\theequation}{\arabic{section}.\arabic{equation}} 

\baselineskip=0.32in


\section{Introduction}

Plasma heating and acceleration are hot topics in the fields of nuclear 
fusion, plasma physics and astrophysics for a long time. A great number of 
heating mechanisms and theories have been proposed. In a collisionless space
plasma environment, wave-particle interactions play significant roles of
plasma heating and acceleration\cite{singh1,chenPOP,WangGRL,Ts,daiPRL,ben}. Even
when collisions are included, plasmas can still be heated by the interplay
between waves and particles to a certain degree\cite{dongPOP,DTp,Leake}. Among a suite of electromagnetic
waves, Alfv\'{e}n waves are generally thought to be the major contributor to
the ion heating and acceleration since they exist pervasively in the solar
wind and interplanetary space\cite{a1,a2,a3,jinApJ}.

In thermodynamic sense, heating of ions by electromagnetic waves will lead
to the dissipation of wave fields, thereby being an irreversible
process. Several recent studies however show that ions can be heated by
turbulent Alfv\'{e}n waves in low-beta plasmas even when no dissipation of
wave fields occurrs\cite%
{wangPRL,wuPRL,wangPOP,yoonPOP,wuPOP1,bwangPOP,SAPL,YN}; a process that
is called ``\emph{pseudoheating}'' or nonresonant wave-particle interaction. 
Various methods have been used to validate this heating mechanism such as
the test-particle approach with analytic solutions\cite{wangPRL,wangPOP}, and
quasi-linear theories\cite{wuPRL,SAPL}. In fact, the name \textquotedblleft
pseudoheating\textquotedblright\ may not be the best term to describe this
``heating'' process due to the kinetic effects of wave spectra\cite%
{wuPRL,yoonPOP}. However, we still use this appellation for consistency in
this paper. The pseudoheating is caused by the wave forces (or their 
spectra) that result in a deformation of the distribution function 
with respect to its initial Maxwellian shape. The mean-square velocity 
fluctuation due to the wave activity leads to an effective broadening 
of the distribution function similar to real heating and thus could mimic 
the genuine heating process\cite{DV}. It is important to point out that the 
heating shown by Ref.\cite{wangPRL} contains both
pseudoheating and genuine heating, which indicates that even in the 
particles' mean-velocity frame, the random kinetic energy of
particles still increases via wave-particle interactions. The real
heating (or the irreversible dissipation of the wave fields) is caused by the
\emph{initial} pitch-angle scattering of newly created ions\cite{bwangPOP},
indicating that there exists the dissipation of wave fields. In contrast, the first adiabatic 
invariant (magnetic moment $\mu=w_{\perp}/B$) remains constant during
the pseudoheating and thus it is an reversible process.
The real heating in Refs.\cite{bwangPOP,wangPRL} is caused by the violation
of the first adiabatic invariant due to the abrupt spatial change of the
magnetic field. It is noteworthy that there is a factor of two difference
between the temperature expressions in Ref.\cite{wangPRL} and Ref.\cite%
{wuPRL}, resulting from the fact that quasilinear theory cannot solve the problem
of the wave damping appropriately. In addition, Wang \emph{et al.} pointed out 
that pitch-angle scattering plays a key role in the pseudoheating process\cite%
{wangPOP}.

In this paper, the heating process we focus on is restricted to the pseudoheating 
and thus the real heating is excluded. We demonstrate that the low-frequency Alfv%
\'{e}n wave propagating along the background magnetic field ${\mathbf{B}}%
_{0}={B}_{0}{\mathbf{i}}_{z}$ can \textquotedblleft \emph{heat}\textquotedblright\
ions. The pseudoheating can be explained either by $E\times B$ drift (in the electric field
of the Alfv\'{e}n wave and the ambient magnetic field $B_0$)
proposed in this paper or the pitch-angle scattering previously investigated\cite{wangPOP}. 
Our analytic results, as will be shown below, are identical as those derived from 
quasilinear theory\cite{wuPRL}.
As shown in previous work\cite{wuPRL,wangPOP}:
\begin{equation}
T_{p\perp }\simeq T_{0}+\frac{W_{B}}{n_{p}}=T_{0}\left( 1+\frac{1}{{\beta }%
_{p}}\frac{B_{W}^{2}}{B_{0}^{2}}\right) \label{t_ps}
\end{equation}%
where $W_{B}=B_{W}^{2}/2\mu _{0}$ and $n_{p}$ are the wave magnetic field
energy density and proton number density, respectively. ${\beta }_{p}$
denotes proton $\beta $ that equals to $\left\langle v\right\rangle
^{2}/v_{A}^{2}$ ($\left\langle v\right\rangle $: the thermal
speed). $B_{W}^{2}/B_{0}^{2}$ represents the ratio of wave-field
energy density to that of the ambient field and $T_{0}=m_{p}\left\langle
v\right\rangle ^{2}/2k_{B}$ is the initial proton temperature. Here $T_{p}$ represents the
``\emph{apparent temperature}'' that should be distinguished from the temperature associated with 
a real heating process\cite{wangPOP}. 

The structure of the remainder of this paper is as follows: We derive and discuss
the analytic results of pseudoheating based on $E \times B$ drift in Section 2. 
In Section 3, test particle simulation results are presented and discussed based on 
the comparison of a monochromatic Alfv\'{e}n wave and a wave spectrum. We also 
briefly discuss the recent observations of large amplitude magnetic perturbations 
associated with Alfv\'{e}n waves and their correlation to the pseudoheating. In the last 
section, conclusions are summarized.

\section{Analytic Theory of Pseudoheating}
Without loss of generality, left-hand circular polarized Alfv\'{e}n waves are 
considered in this paper. The wave magnetic field vector can thus be
expressed as
\begin{equation}
\mathbf{B}_{W}=\sum_{k}{B}_{k}(\cos {\phi }_{k}{\mathbf{i}}_{x}-\sin {\phi }%
_{k}{\mathbf{i}}_{y}),
\end{equation}%
From the Faraday's law and the dispersion relation (Amp\`{e}re's  law) $\omega=kv_A$, 
the electric field vector can be written as
\begin{equation}
\mathbf{E}_{W}=-v_{A}{\mathbf{i}}_{z}\times \mathbf{B}_{W}  \label{efv}
\end{equation}%
where ${\mathbf{i}}_{x}$, ${\mathbf{i}}_{y}$ and ${\mathbf{i}}_{z}$ are unit
directional vectors, ${\phi }_{k}=k({v}_{A}t-z)+\varphi _{k}$ denotes the
wave phase, $\varphi _{k}$ is the random phase for mode $k$ and $v_{A}=B_{0}/%
\sqrt{\mu _{0}n_{p}m_{p}}$ represents the Alfv\'{e}n speed. 
According to the linear approximation, we use Eq.(\ref{efv}) and the first
order term of the generalized Ohmic law $\mathbf{E}_{W} = -\mathbf v_{\perp} \times \mathbf B_0$ to derive the $E\times B$ drift velocity, $\mathbf{v%
}_{E}$, that can be expressed as follows:
\begin{equation}
\mathbf{v}_{E}=\mathbf{v}_{\perp }=\frac{\mathbf{E}_{W}\times \mathbf{B}_{0}%
}{B_{0}^{2}}=\frac{-v_{A}{\mathbf{i}}_{z}\times \mathbf{B}_{W}\times \mathbf{%
B}_{0}}{B_{0}^{2}} \label{vdrift}
\end{equation}
The derivation above is consistent with the traditional procedure to derive the classical 
Alfv\'{e}n wave solution\cite{DV} in the ideal magnetohydrodynamic (MHD) system.
It indicates that the randomized proton motion is
actually parasitic to wave fields due to the fact that
the drift velocity $\mathbf{v}_{E}$ is expected to disappear if the waves subside. Given the
assumption that the characteristic spatial scale of the system is much
larger than typical Alfv\'{e}n wavelength\cite{wangPRL}, the energy density
associated with the wave magnetic field can be written as:
\begin{equation}
W_{B}=\frac{1}{2}n_{p}m_{p}\left\langle \left( \mathbf{v}_{E}-\left\langle
\mathbf{v}_{E}\right\rangle \right) ^{2}\right\rangle =\frac{1}{2}%
n_{p}m_{p}v_{A}^{2}\frac{{B}_{W}^{2}}{B_{0}^{2}}=\frac{1}{2}n_{p}m_{p}\frac{%
B_{0}^{2}}{\mu _{0}n_{p}m_{p}}\frac{{B}_{W}^{2}}{B_{0}^{2}}=\frac{B_{W}^{2}}{%
2\mu _{0}} \label{WB}
\end{equation}
where the bracket $\left\langle \cdot \right\rangle $\ denotes an average
over all particles. Given the large characteristic spatial scale of the system, $l$, 
as described above, the following approximation is valid: 
\begin{equation}
\left\langle \mathbf{v}_{E}\right\rangle=\lim_{l\rightarrow \infty}\frac{1}{l} \int_{0}^{l} \mathbf{v}_{E} dz =0
\end{equation}
The temperature expression shown below is the same as
Eq.(\ref{t_ps}),
\begin{equation}
T_{p\perp }=T_{0}+\frac{1}{2}n_{p}m_{p}\left\langle \left( \mathbf{v}%
_{E}-\left\langle \mathbf{v}_{E}\right\rangle \right) ^{2}\right\rangle
=T_{0}+\frac{B_{W}^{2}}{2\mu _{0}}=T_{0}\left( 1+\frac{1}{\beta _{p}}\frac{%
B_{W}^{2}}{B_{0}^{2}}\right)  \label{temp}
\end{equation}
The results shown in Eqs.(\ref{WB})\&(\ref{temp}) are consistent with the MHD theory due to the
fact that the energy density associated with the Alfv\'{e}n wave magnetic field, $B_W^2/2\mu_0$, 
equals to the ion (fluid) kinetic energy density, $B_W^2/2\mu_0=B_0^2u_1^2/(2\mu_0v_A^2) =\rho_0 u_1^2/2$, where $u_1$ is the perturbed ion (fluid) velocity. The consistency between Eqs.(\ref{WB})\&(\ref{temp}) and the MHD theory\cite{YN} clearly 
indicates that the analytic theory described here incorporates the local equilibrium velocity distribution of the ions. In the following section, we adopt the test particle approach to simulate
pesudoheating and the results will be presented and discussed in detail.

\section{Test Particle Simulations of Pseudoheating}

We start with a linearly polarized Alfv\'{e}n wave with wave magnetic field
vector $\mathbf{B}_{W}=\sum_{k}{B}_{k}\cos {\phi }_{k}{\mathbf{i}}_{y}$ in
order to show that the drift velocity is in $E\times B$ direction. This
is the basis of further understanding. Then we conduct two case
studies. In case one, we consider a monochromatic dispersionless Alfv\'{e}n
wave with frequency $\omega $=0.05$\Omega _{p}$. In case two, we test a spectrum of Alfv%
\'{e}n waves with random phase $\varphi _{k}$, and the frequencies of the
wave modes can be calculated as follows: $\omega _{i}=\omega
_{1}+(i-1)\triangle \omega $ ($i$=1,2,..., $N$; $N$=41), where $\triangle
\omega =(\omega _{N}-\omega _{1})/(N-1)$; $\omega _{1}$=0.01$\Omega _{p}$
and $\omega _{N}$=0.05$\Omega _{p}$. The amplitude of each wave
mode (only one wave mode in case one) is considered to be equal but changes
gradually with time such that $B_{W}^{2}=\sum_{k}B_{k}^{2}=\epsilon
(t)B_{0}^{2}$, where

\begin{center}
$\epsilon (t)=\left\{
\begin{array}{ccc}
\epsilon _{0}e^{-(t-t_{1})^{2}/\tau ^{2}}, & if & t<t_{1}, \\
\epsilon _{0}, & if & t_{1}\leq t\leq t_{2}, \\
\epsilon _{0}e^{-(t-t_{2})^{2}/\tau ^{2}}, & if & t>t_{2}.%
\end{array}%
\right. $
\end{center}

\begin{figure}[tbp]
\centering
\includegraphics[scale=0.4]{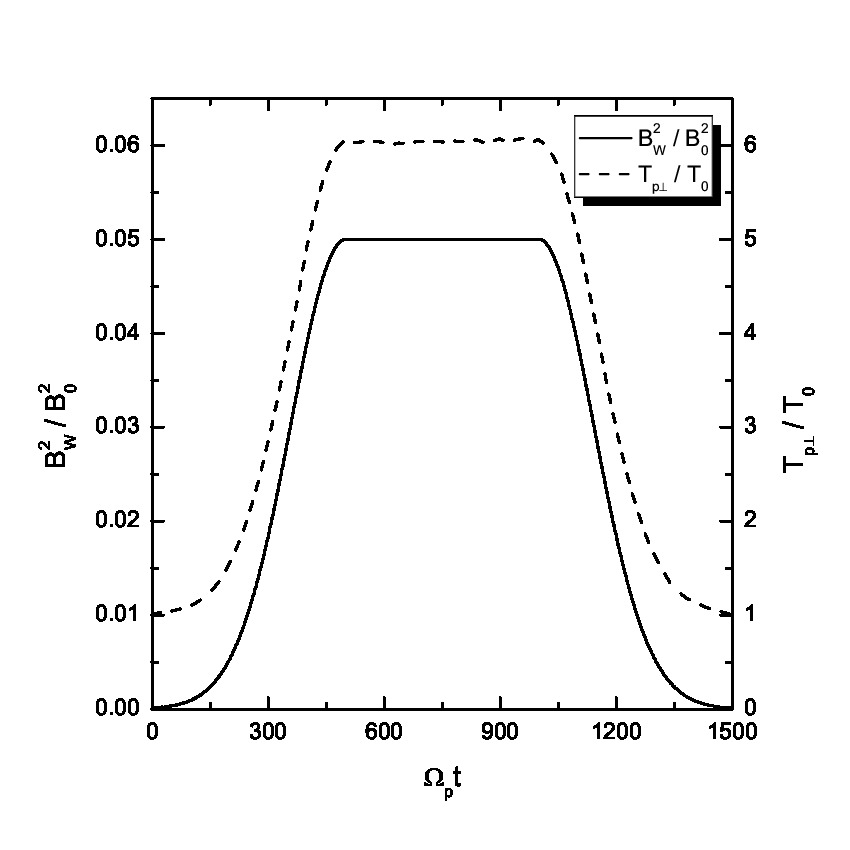}
\caption{Wave field strength $B_W^2/B_0^2$ (solid line) and perpendicular apparent
temperature $T_{p\perp}$ (dashed line) \emph{vs} time; $v_{p}$=0.07$v_{A}$.}
\label{fig1}
\end{figure}

Similar to previous work\cite{wangPOP}, we set $t_{1}$=500$\Omega _{p}^{-1}$%
, $t_{2}$=1000$\Omega _{p}^{-1}$, $\tau $=200$\Omega _{p}^{-1}$, and $%
\epsilon _{0}$=0.05, where $\Omega _{p}$ is the proton gyrofrequency. The
initial velocities of test particles are randomly distributed and possess a
Maxwellian distribution with thermal speed $v_{p}$=0.07$v_{A}$ (the
situations with $v_{p}$=0.01$v_{A}$, 0.03$v_{A}$, and 0.15$v_{A}$ are also
discussed later). The numerical scheme is similar to what described in Ref.%
\cite{wangPRL,dongPOP}. It is noteworthy that the ion thermal speed $v_{p}$
in this paper and in Refs.\cite{dongPOP,wangPRL,wangPOP,bwangPOP} is defined as
$v_{p}=\left( k_{B}T/m\right) ^{1/2}$\cite{book} while the general thermal
speed $\left\langle v\right\rangle =$ $\left( 2k_{B}T/m\right) ^{1/2}$,
causing $v_{p}^{2}=\left\langle v\right\rangle ^{2}/2$. Thus it leads to 
a factor of two difference for proton $\beta _{p}$. The basic reason causing
this difference is the different definition of temperature, i.e., $%
T=m\left\langle v\right\rangle ^{2}/2k_{B}$ or $T=mv_{p}^{2}/k_{B}$. The
one-dimensional Maxwellian velocity distribution based on $\left\langle
v\right\rangle $ and $v_{p}$ can be expressed as follows\cite{book}:
\begin{equation}
f_{v}\left( v_{i=x,y,z}\right) =\frac{n}{\left( \pi \left\langle
v_{i}\right\rangle ^{2}\right) ^{1/2}}\exp \left( -\frac{v_{i}^{2}}{%
\left\langle v_{i}\right\rangle ^{2}}\right) =\frac{n}{\left( 2\pi
v_{p}^{2}\right) ^{1/2}}\exp \left( -\frac{v_{i}^{2}}{2v_{p}^{2}}\right)
\label{MaxDis}
\end{equation}%
Different definitions of thermal speed or temperature, however, do not
affect the final results since self-consistent definition is maintained
throughout the previous studies\cite{dongPOP,wangPRL,wangPOP,bwangPOP,wuPRL}.

\begin{figure}[tbp]
\centering
\includegraphics[scale=0.4]{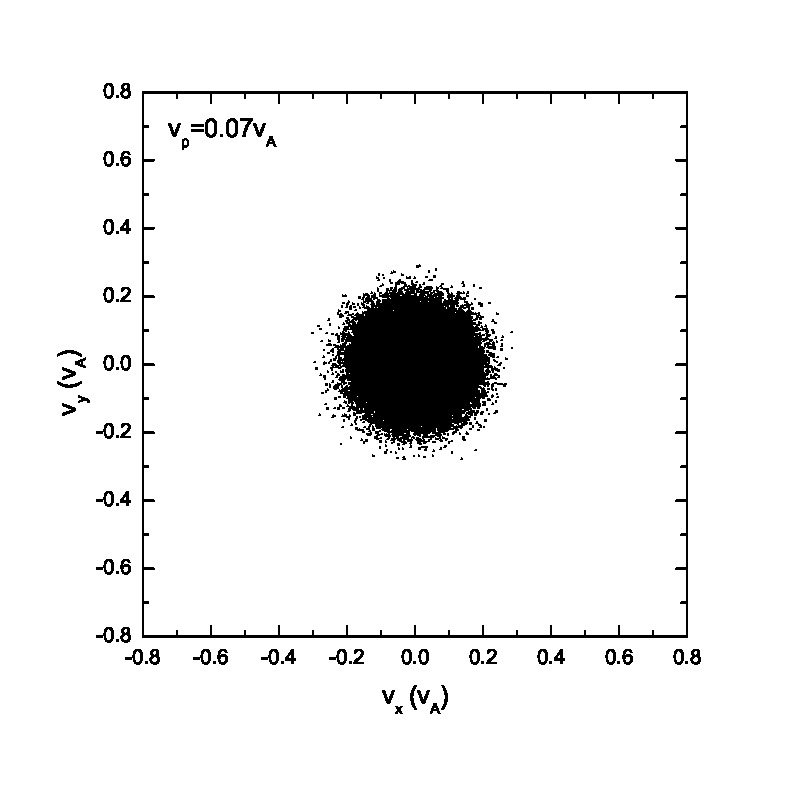} %
\includegraphics[scale=0.4]{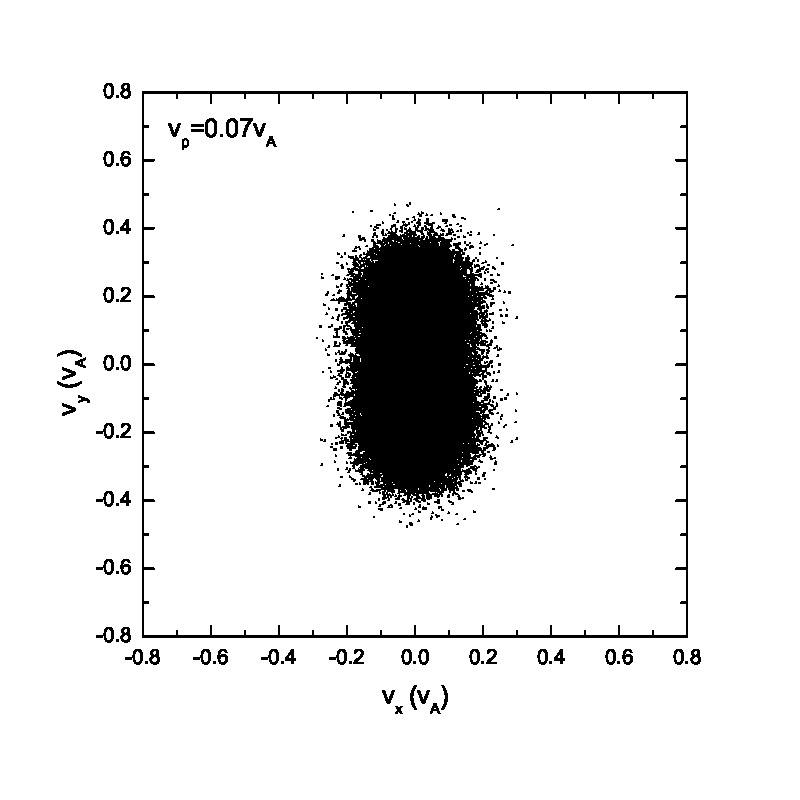}
\caption{Velocity scatter plots of test particles in the $v_x-v_y$ space
for a linearly polarized Alfv\'{e}n wave $\mathbf{B}_{W}=\sum_{k}{B}_{k}\cos
{\protect\phi }_{k}{\mathbf{i}}_{y}$ at $\Omega _{p}$t=0 (left) and $\Omega
_{p}$t=600 (right); $v_{p}$=0.07$v_{A}$.}
\label{fig11}
\end{figure}

Fig.\ref{fig1} shows the dependence of apparent temperature ($T_{p\perp }$)
on time-dependent wave field strength $B_{W}^{2}/B_{0}^{2}$ under a spectrum
of Alfv\'{e}n waves. The result of a monochromatic dispersionless Alfv\'{e}%
n wave is almost the same as that under a spectrum of Alfv\'{e}n waves,
which is in consistency with the analytic result shown above; the
temperature expression Eq.(\ref{temp}) is independent of the number of wave
modes, $N$. In Fig.\ref{fig1}, apparent temperature versus time step shows
the same tendency as the wave field strength, indicating that the proton 
temperature returns to its original value when the waves subside, in
accordance with the work of Wang \emph{et al.}\cite{wangPOP,bwangPOP}. It implies 
once again that the pseudoheating process is parasitic to the waves, as indicated 
by the aforementioned $E\times B$ drift. It is very important that the 
consistency between Eqs.(\ref{WB})\&(\ref{temp}) and the numerical solution shown
in Fig.\ref{fig1} indirectly ensures the self-consistency of the test particle
simulation debated in the Refs.\cite{Dongr,Luc}. Furthermore, this process 
is analogous to the ion motion in a magnetic mirror, where the magnetic moment 
is invariant. This physical picture helps us to better understand  the reversibility in 
pseudoheating.
\begin{figure}[tbp]
\centering
\includegraphics[scale=0.35]{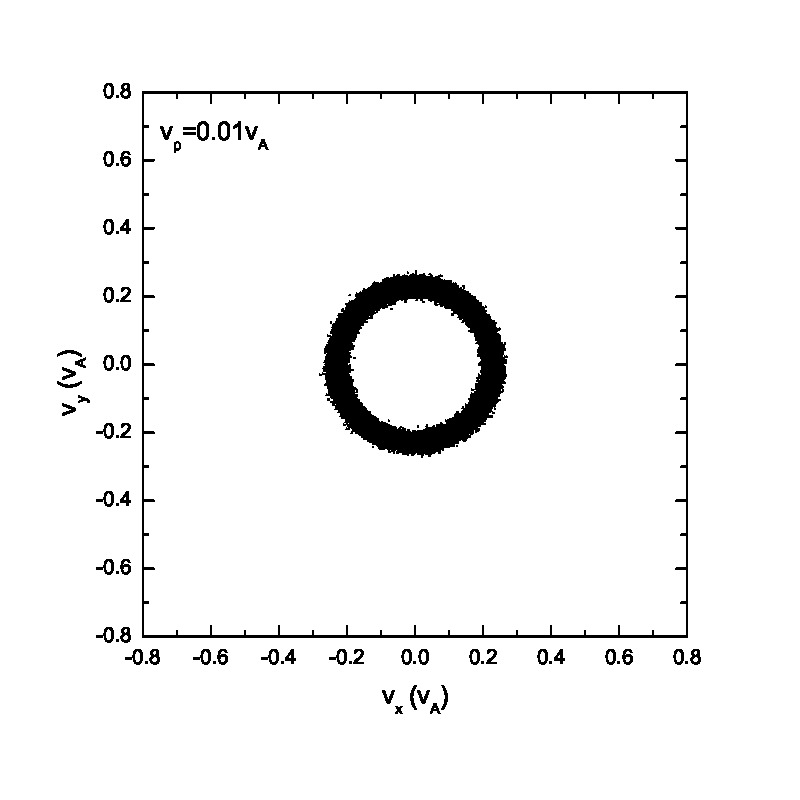} %
\includegraphics[scale=0.35]{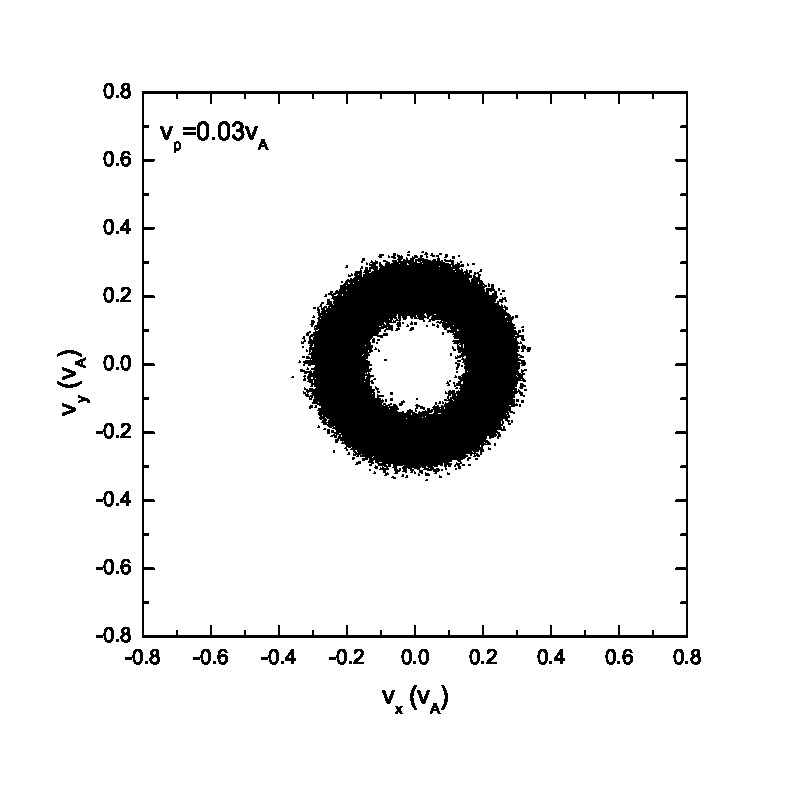} %
\includegraphics[scale=0.35]{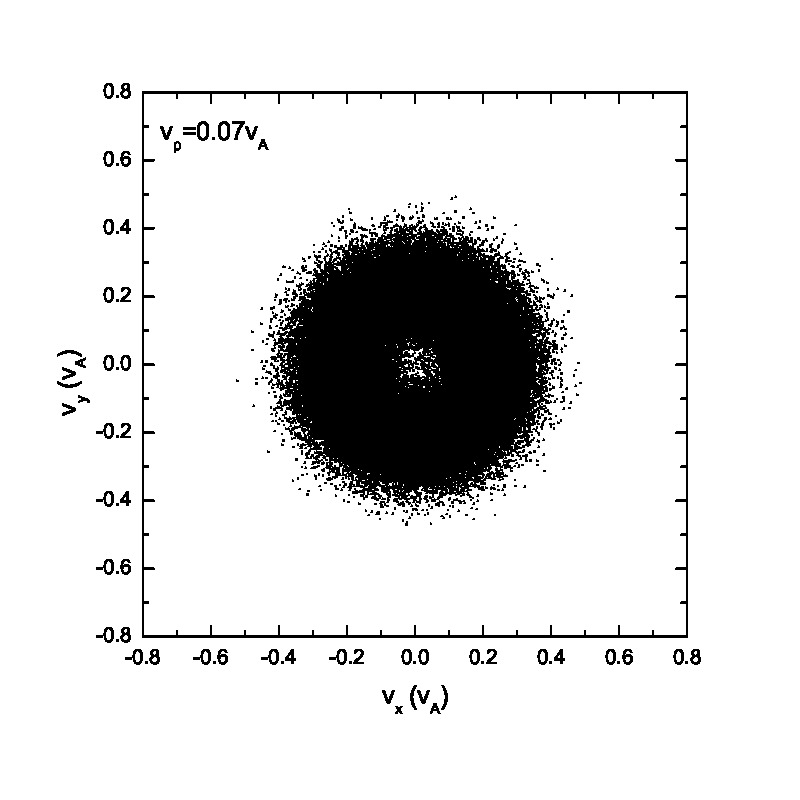} %
\includegraphics[scale=0.35]{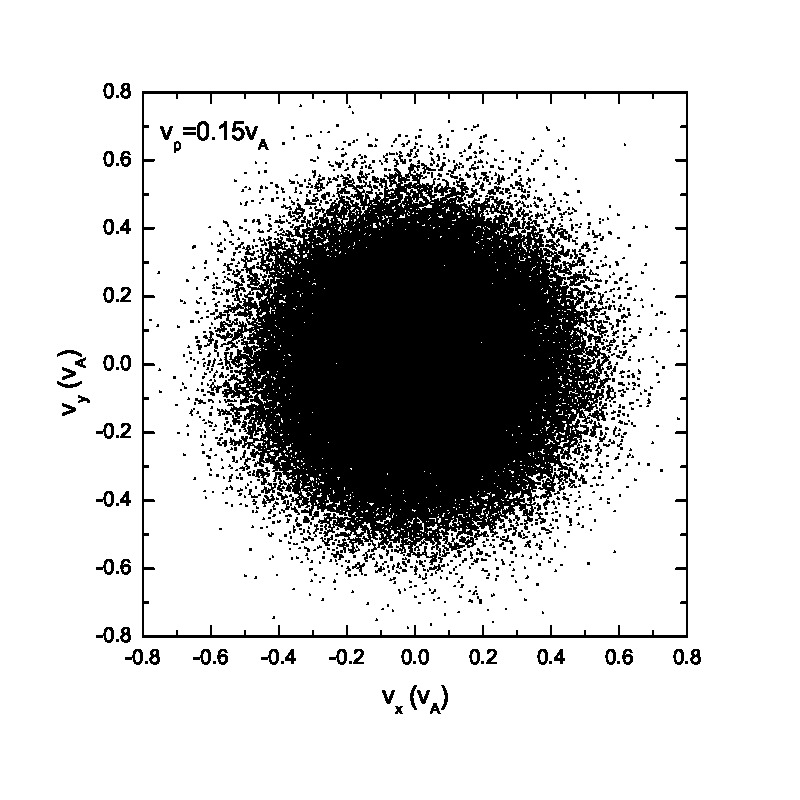}
\caption{ Velocity scatter plots of test particles in the $v_{x}-v_{y}$
space for a circularly polarized Alfv\'{e}n wave at $\Omega _{p}$t=600. (a) $v_{p}$=0.01$v_{A}$
(b) $v_{p}$=0.03$v_{A}$, (c) $v_{p}$=0.07$v_{A}$, and (d) $v_{p}$=0.15$v_{A}$.}
\label{fig31}
\end{figure}

In order to show that the pesudoheating is caused by $E\times B$ drift,
we first illustrate the velocity scatter plots of test particles in the
$v_x-v_y$ space for a linearly polarized Alfv\'{e}n wave. According to Eq.(\ref{efv}%
), if $\mathbf{B}_{W}$ is in the $y$ direction, $\mathbf{E}_{W}$ is in
the $x$ direction; therefore, $\mathbf{E}_{W}\times \mathbf{B}_{0}$ is in
the $y$ direction (ignore the negative sign here). As indicated in 
Fig.\ref{fig11}, the drift velocity is in the $y$ direction, in agreement with 
the analytic results. To be consistent with the previous work\cite
{wangPRL,wuPRL,wangPOP,yoonPOP,wuPOP1,bwangPOP,SAPL,YN}, we will
focus on the circularly polarized condition in the following paragraphs. The
further discussions are based on the comparison of a monochromatic Alfv\'{e}n 
wave and a wave spectrum.

\begin{figure}[tbp]
\centering
\includegraphics[scale=0.2]{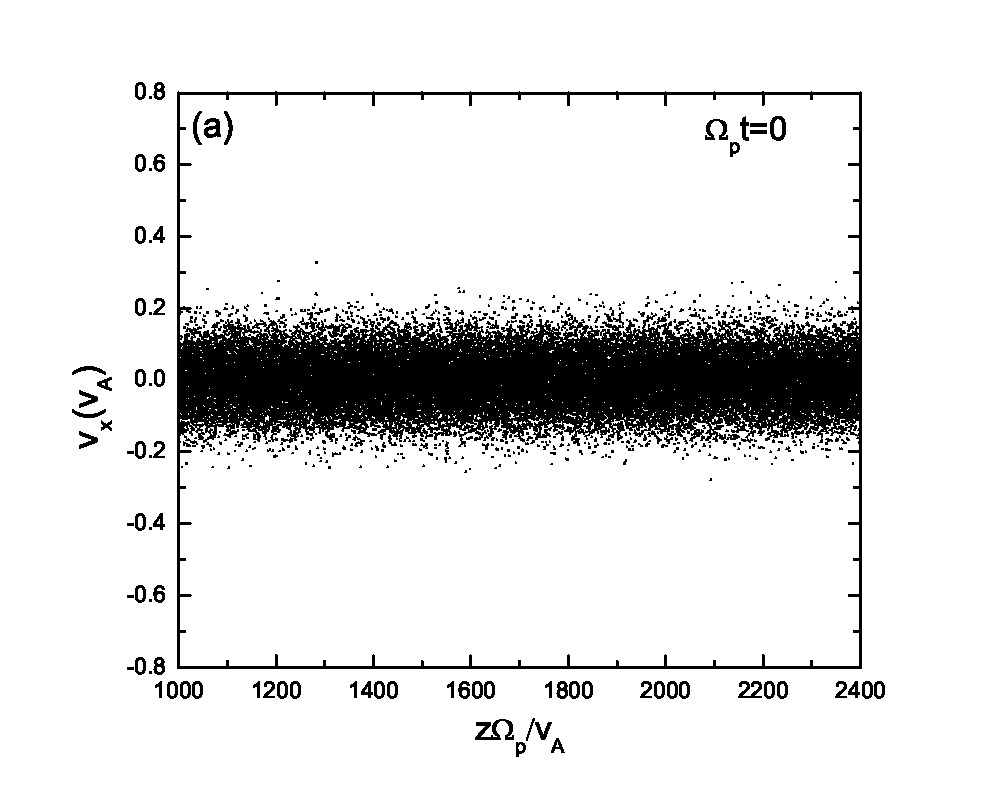} 
\includegraphics[scale=0.2]{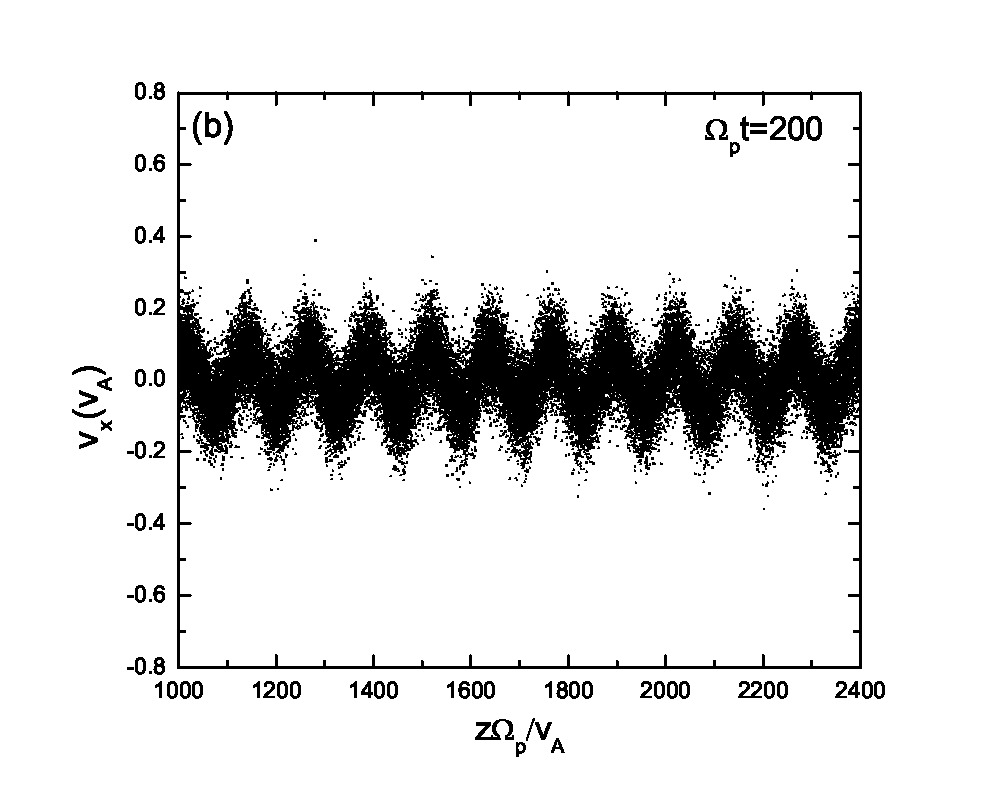}
\includegraphics[scale=0.2]{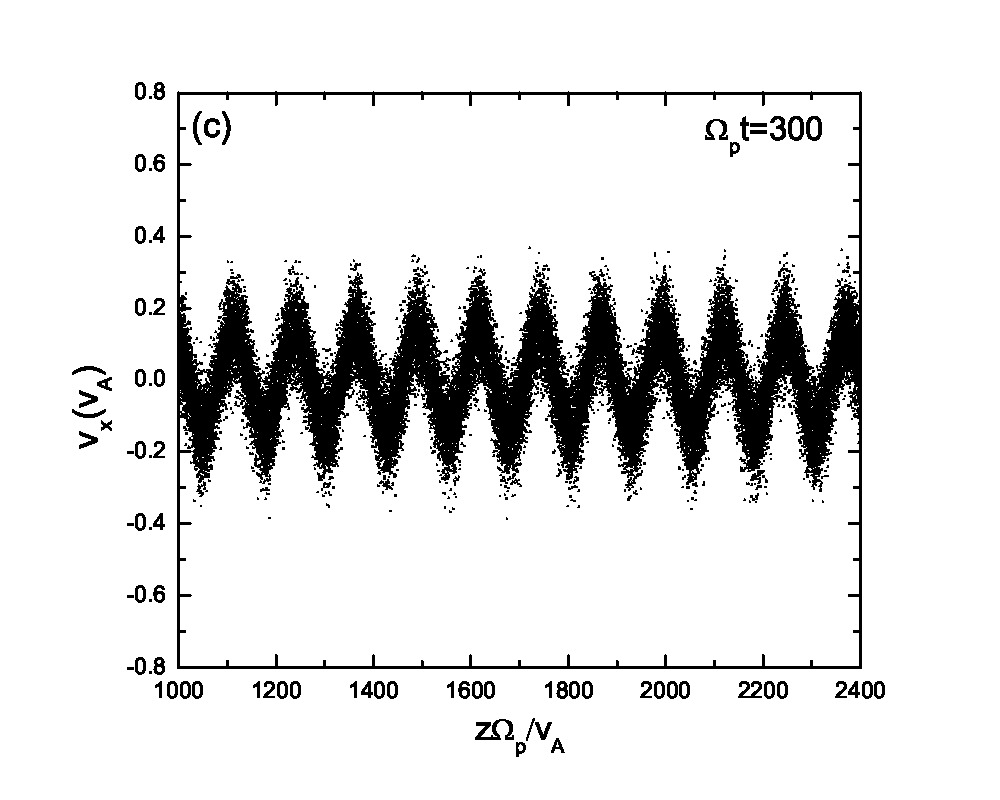} %
\includegraphics[scale=0.2]{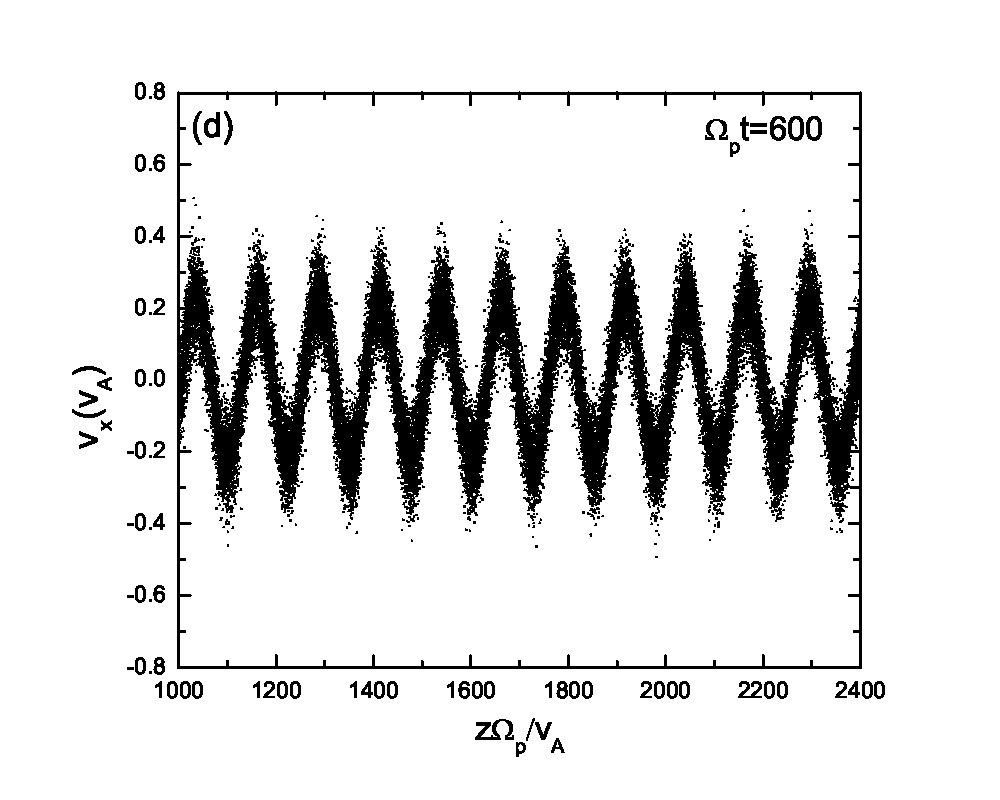} %
\includegraphics[scale=0.2]{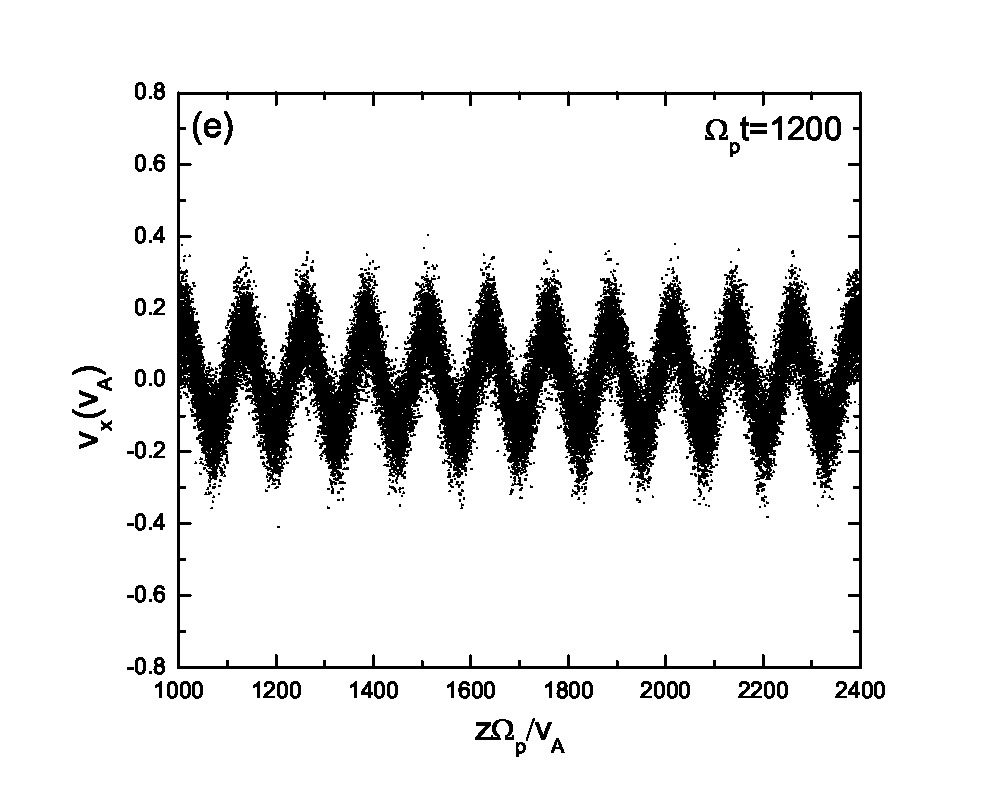} %
\includegraphics[scale=0.2]{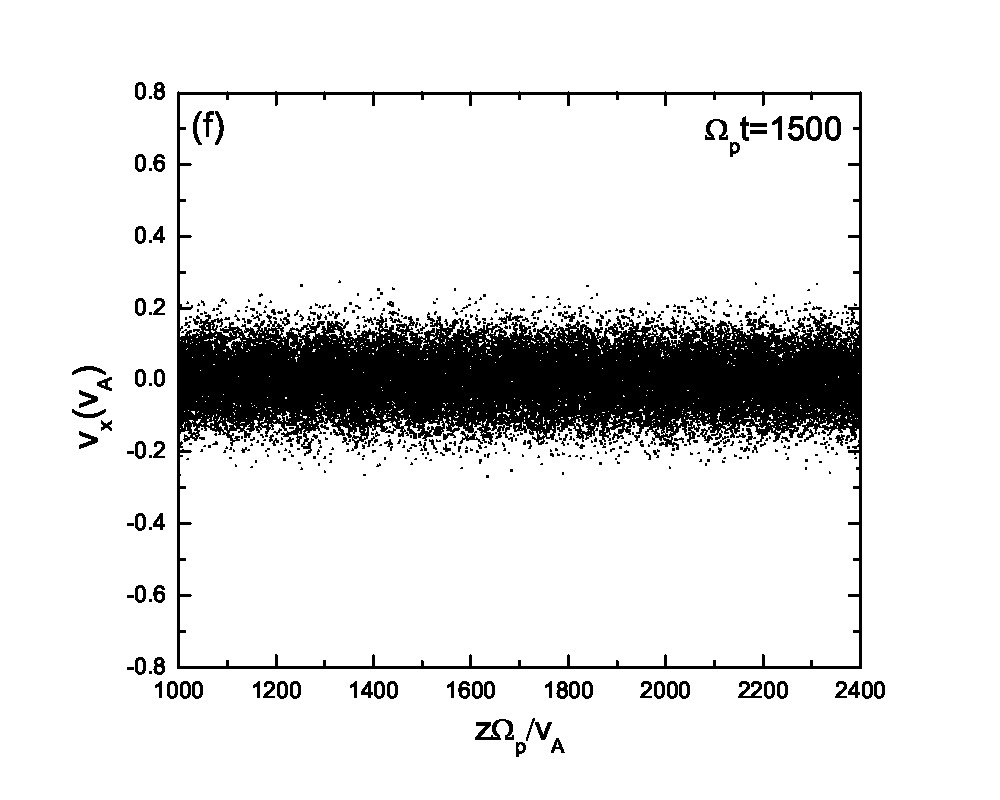}
\caption{Scatter plots of protons between 1000 and 2400$v_{A}\Omega
_{p}^{-1} $ at different times, $\Omega _{p}$t=0, 200, 300, 600, 1200 and
1500 in the $v_{x}$-$z$ phase space for a circularly polarized Alfv\'{e}n
wave; $v_{p}$=0.07$v_{A}$.}
\label{fig3}
\end{figure}

Fig.\ref{fig31} presents the velocity scatter plots of ions for a circularly
polarized Alfv\'{e}n wave with different initial thermal speeds: 
$v_{p}$=0.01$v_{A}$, 0.03$v_{A}$, 0.07$v_{A}$ and 0.15$v_{A}$.
Inspection of Fig.\ref{fig31} reveals that when the thermal speed is quite
small with respect to the Alfv\'{e}n speed $v_{A}$ 
(i.e., $v_{p}$=0.01$v_{A}$ and 0.03$v_{A}$), test particles form a
ring distribution in the $v_{x}-v_{y}$ velocity space. Although the ring 
distribution eventually can be filled with test particles with
different velocities when $v_{p}$ becomes large, the heating efficiency
becomes relatively low [refer to Eq.(\ref{temp})]. Besides the velocity scatter 
plot, particle distribution in the phase space is also essential for our
understanding of the pesudoheating. Fig.\ref{fig3} shows the scatter plots 
of protons between 1000 and 2400$ v_{A}\Omega _{p}^{-1}$ at
different times $\Omega _{p}$t=0, 200, 300, 600, 1200 and 1500 for a 
monochromatic dispersionless Alfv\'{e}n wave. The results agree 
with that shown in Fig.\ref{fig1}; the stronger the wave fields are, the more 
obvious the velocity fluctuations are. A common feature that 
stands out in Fig.\ref{fig3} is that the particle motion is periodic 
under a monochromatic Alfv\'{e}n wave. It indicates that the kinetic 
behavior of test particles under a monochromatic dispersionless wave versus
a wave spectrum, as will be shown below, is fairly different.

\begin{figure}[tbp]
\centering
\includegraphics[scale=0.32]{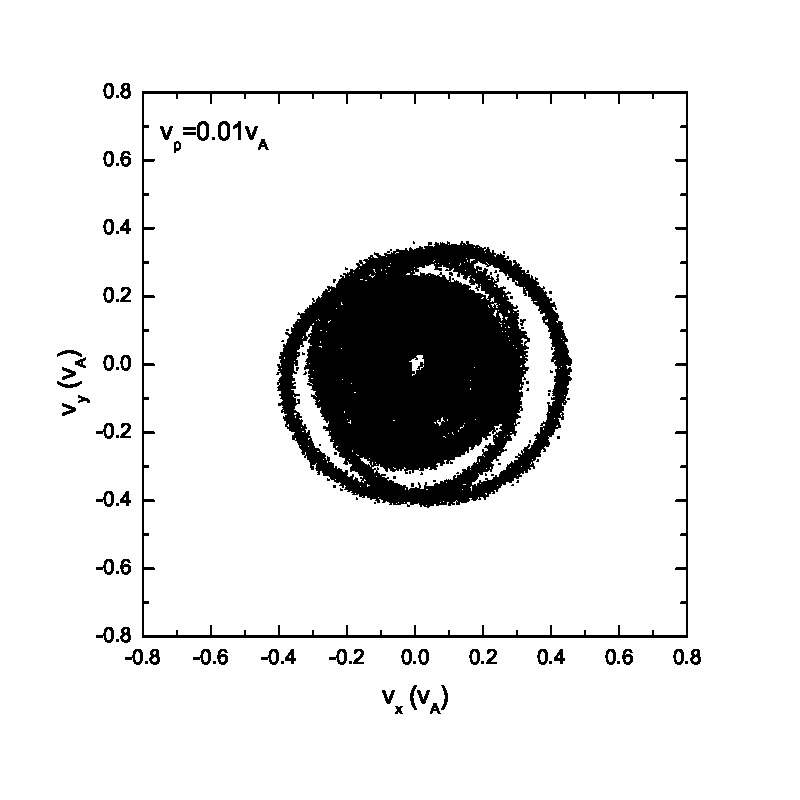} %
\includegraphics[scale=0.32]{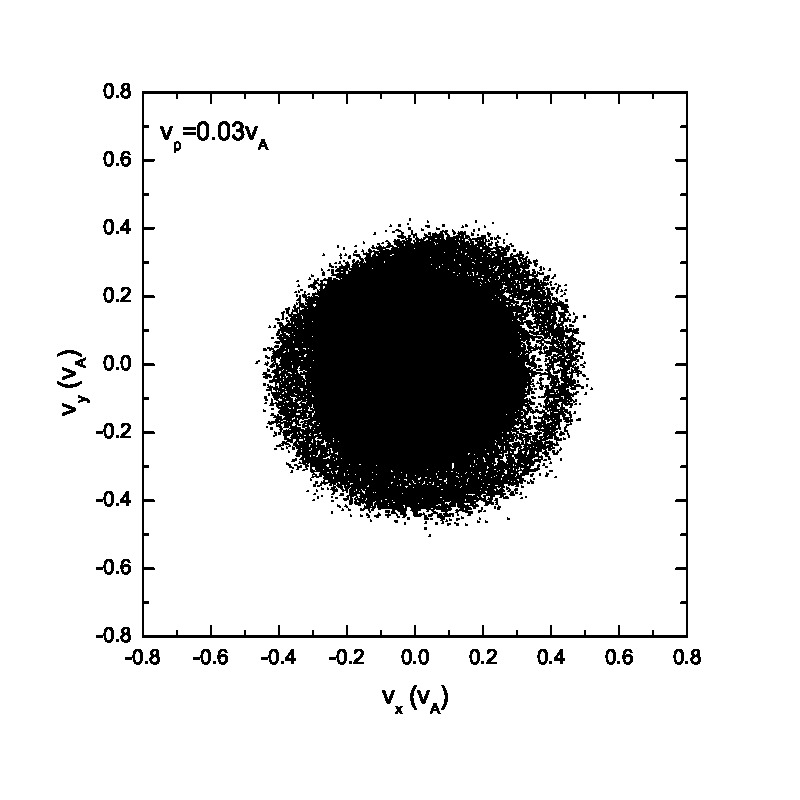} %
\includegraphics[scale=0.32]{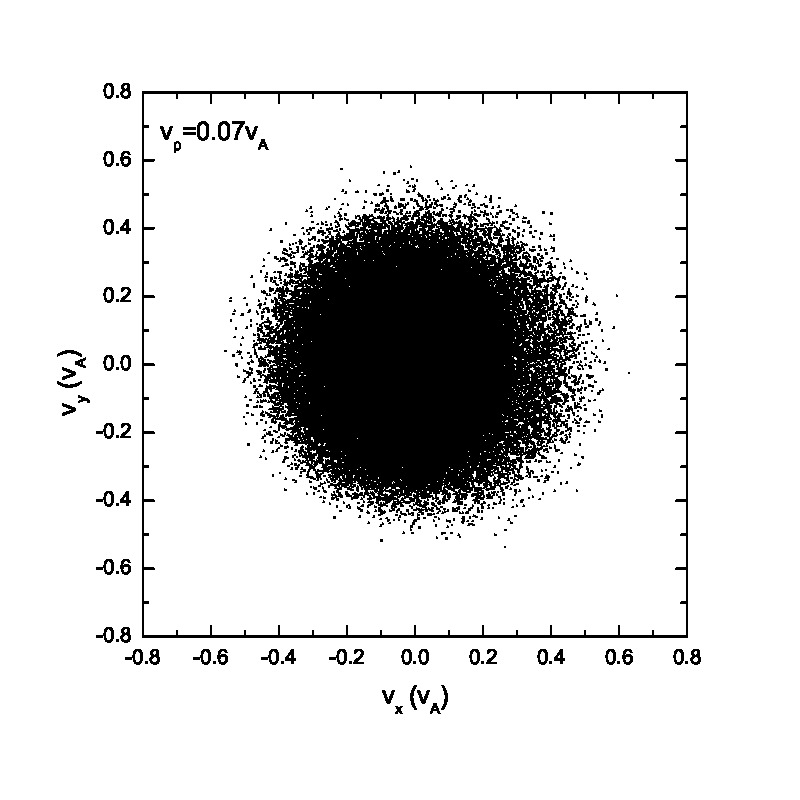} %
\includegraphics[scale=0.32]{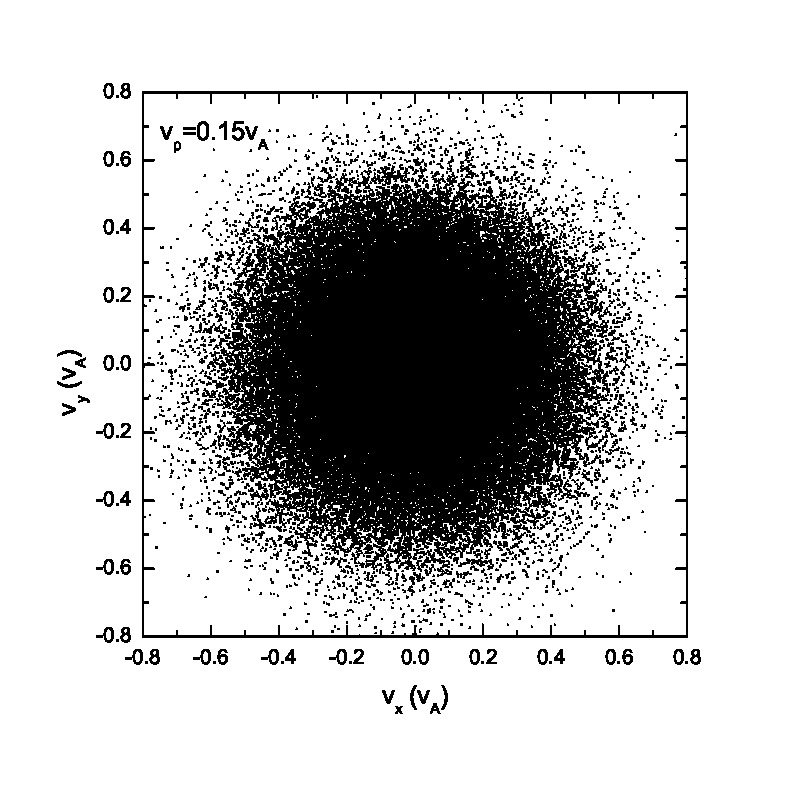} %
\includegraphics[scale=0.32]{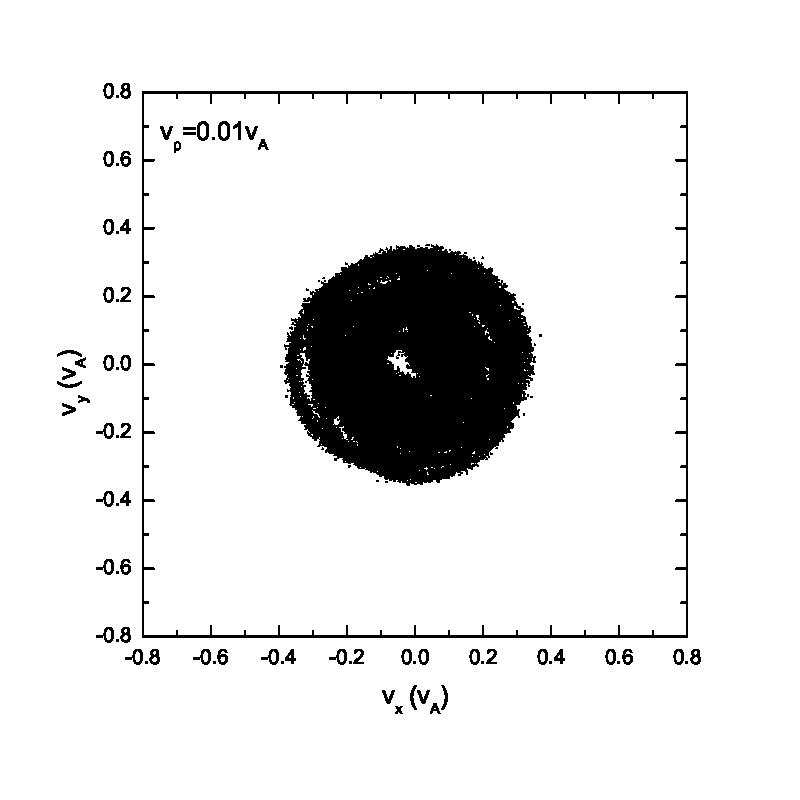} %
\includegraphics[scale=0.32]{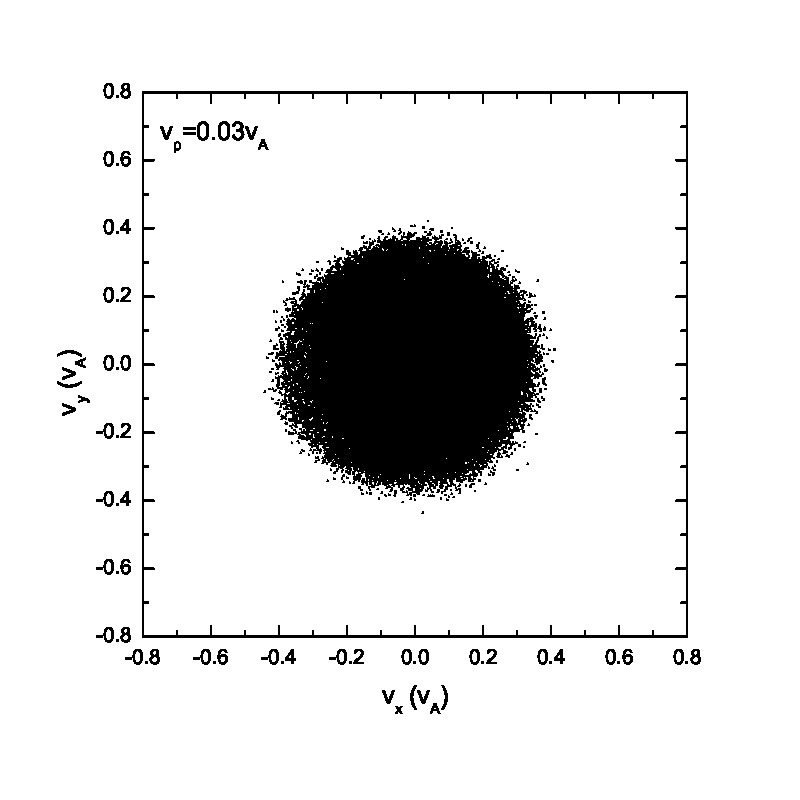}
\caption{Velocity scatter plots of test particles in the $v_{x}-v_{y}$
space for a spectrum of circularly polarized Alfv\'{e}n waves with random
phases $\protect\varphi _{k}$ at $\Omega _{p}$t=600. (a) $v_{p}$=0.01$v_{A}$, (b) $v_{p}$=0.03$%
v_{A}$, (c) $v_{p}$=0.07$v_{A}$, (d) $v_{p}$=0.15$v_{A}$, (e) $v_{p}$=0.01$%
v_{A}$, and (f) $v_{p}$=0.03$v_{A}$; (a)-(d): 41 wave modes, (e)-(f): 1001
wave modes.}
\label{fig21}
\end{figure}

\begin{figure}[tbp]
\centering
\includegraphics[scale=0.2]{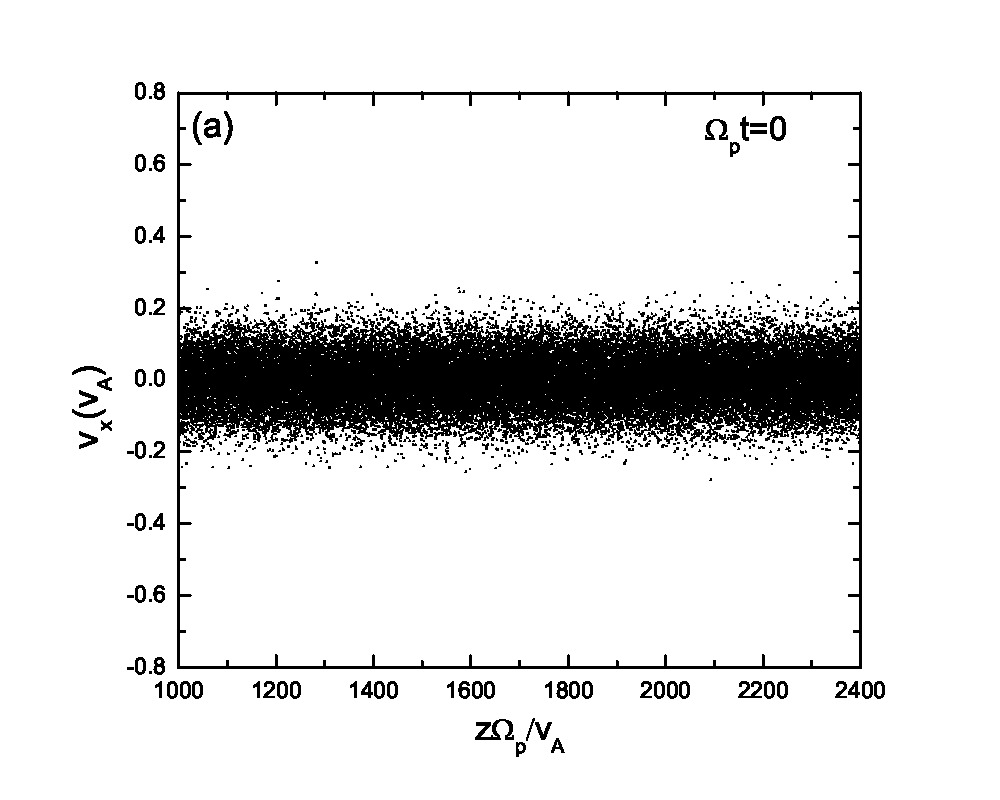} 
\includegraphics[scale=0.2]{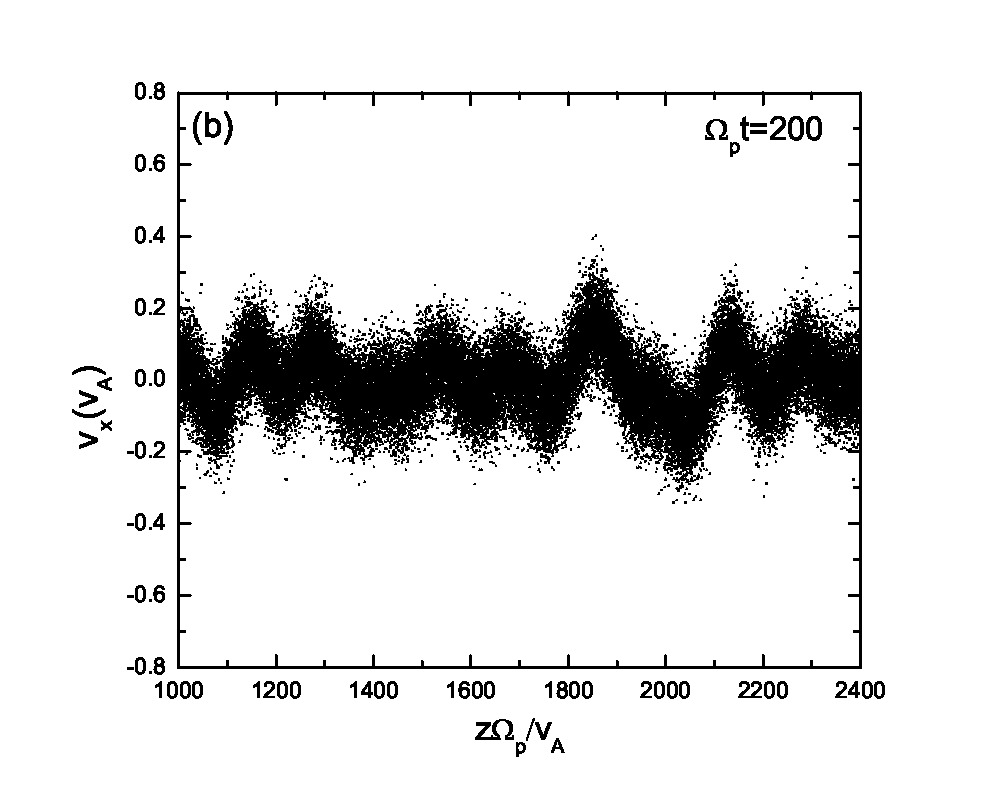}
\includegraphics[scale=0.2]{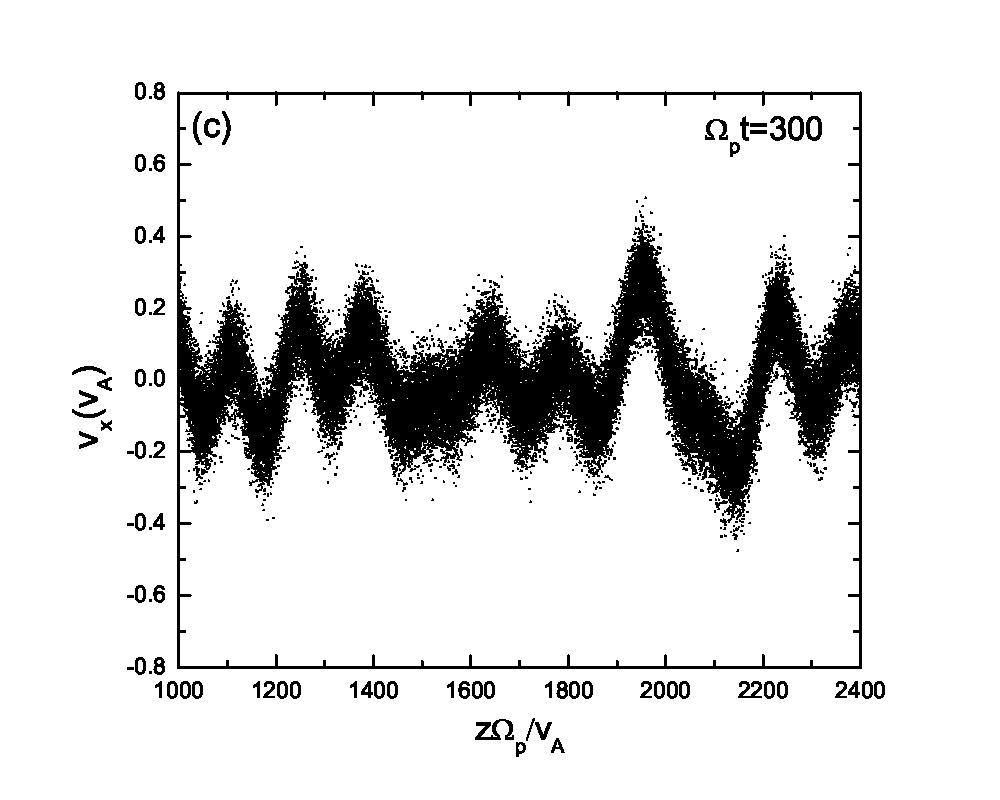} %
\includegraphics[scale=0.2]{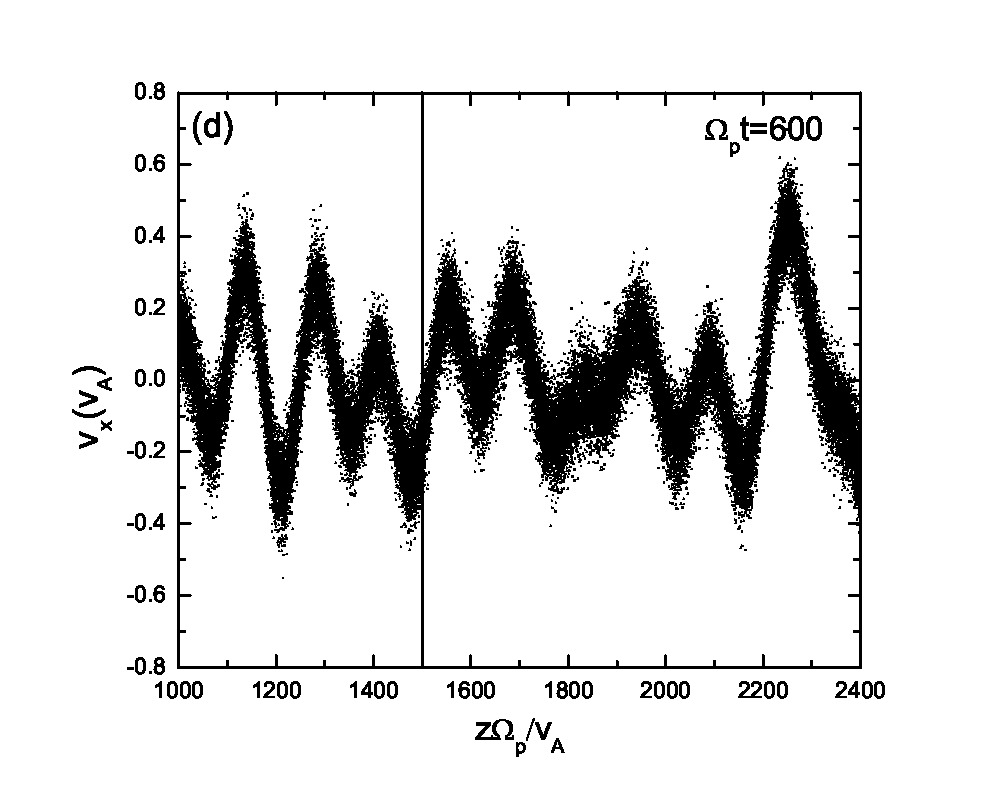} %
\includegraphics[scale=0.2]{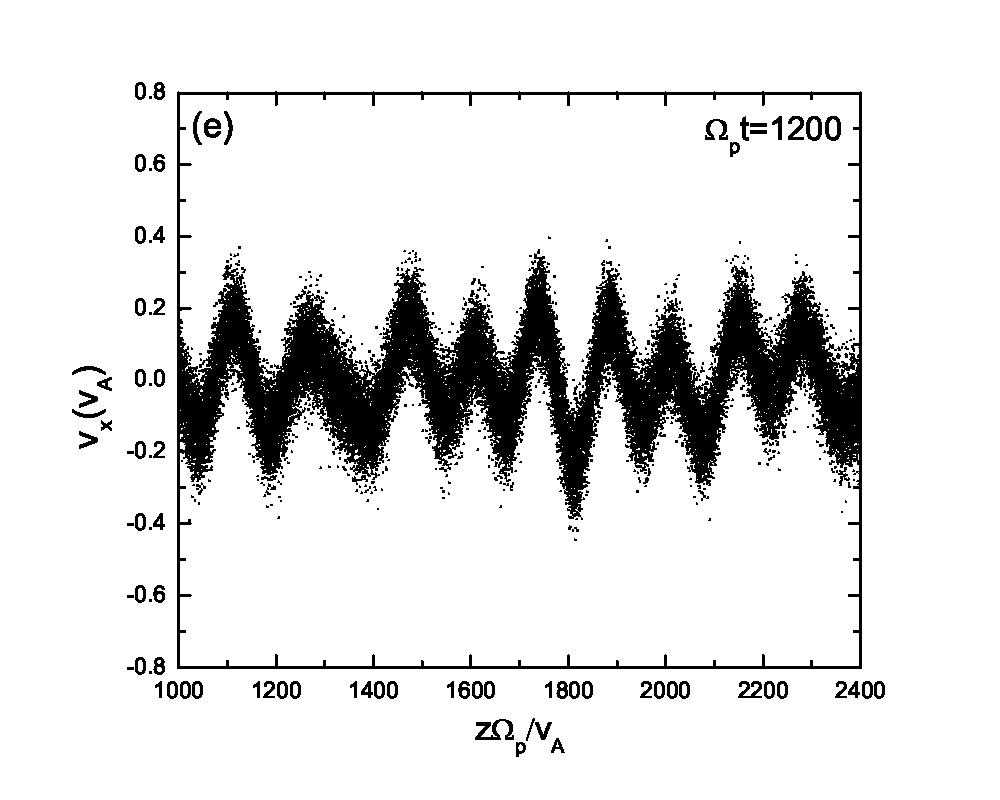} %
\includegraphics[scale=0.2]{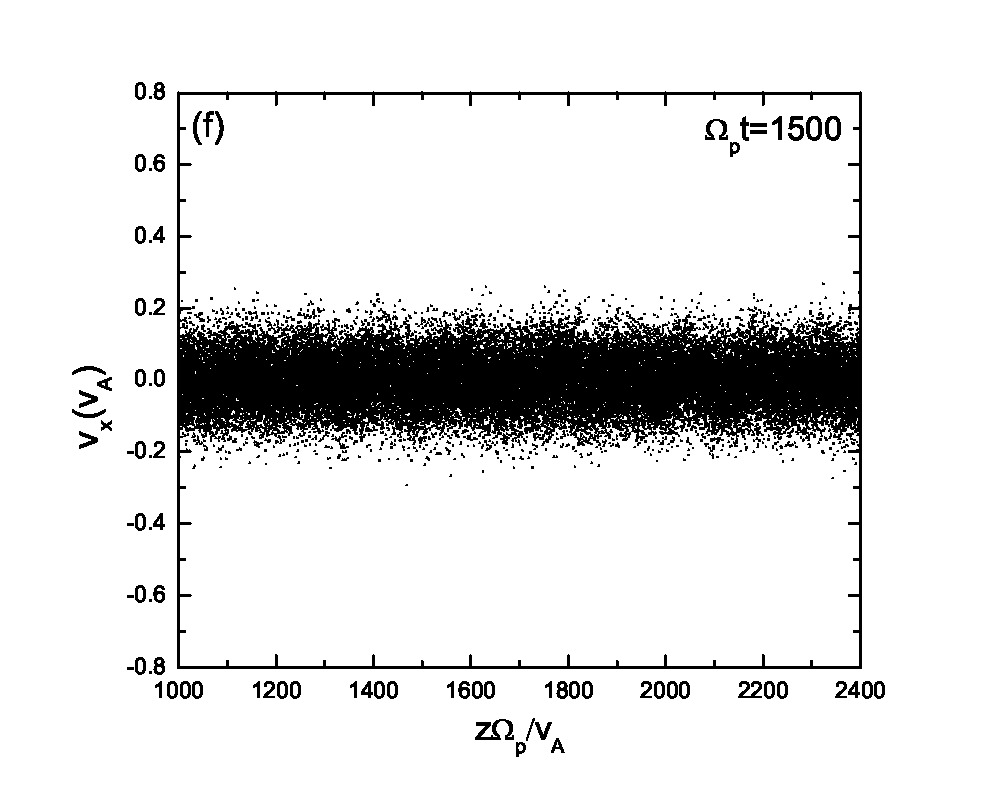}
\caption{Scatter plots of protons between 1000 and 2400$v_{A}\Omega
_{p}^{-1} $ at different times, $\Omega _{p}$t=0, 200, 300, 600, 1200 and
1500 in the $v_{x}$-$z$ phase space for an Alfv\'{e}n wave spectrum with
random phases $\protect\varphi _{k}$; $v_{p}$=0.07$v_{A}$.}
\label{fig2}
\end{figure}

\begin{figure}[tbp]
\centering
\includegraphics[scale=0.25]{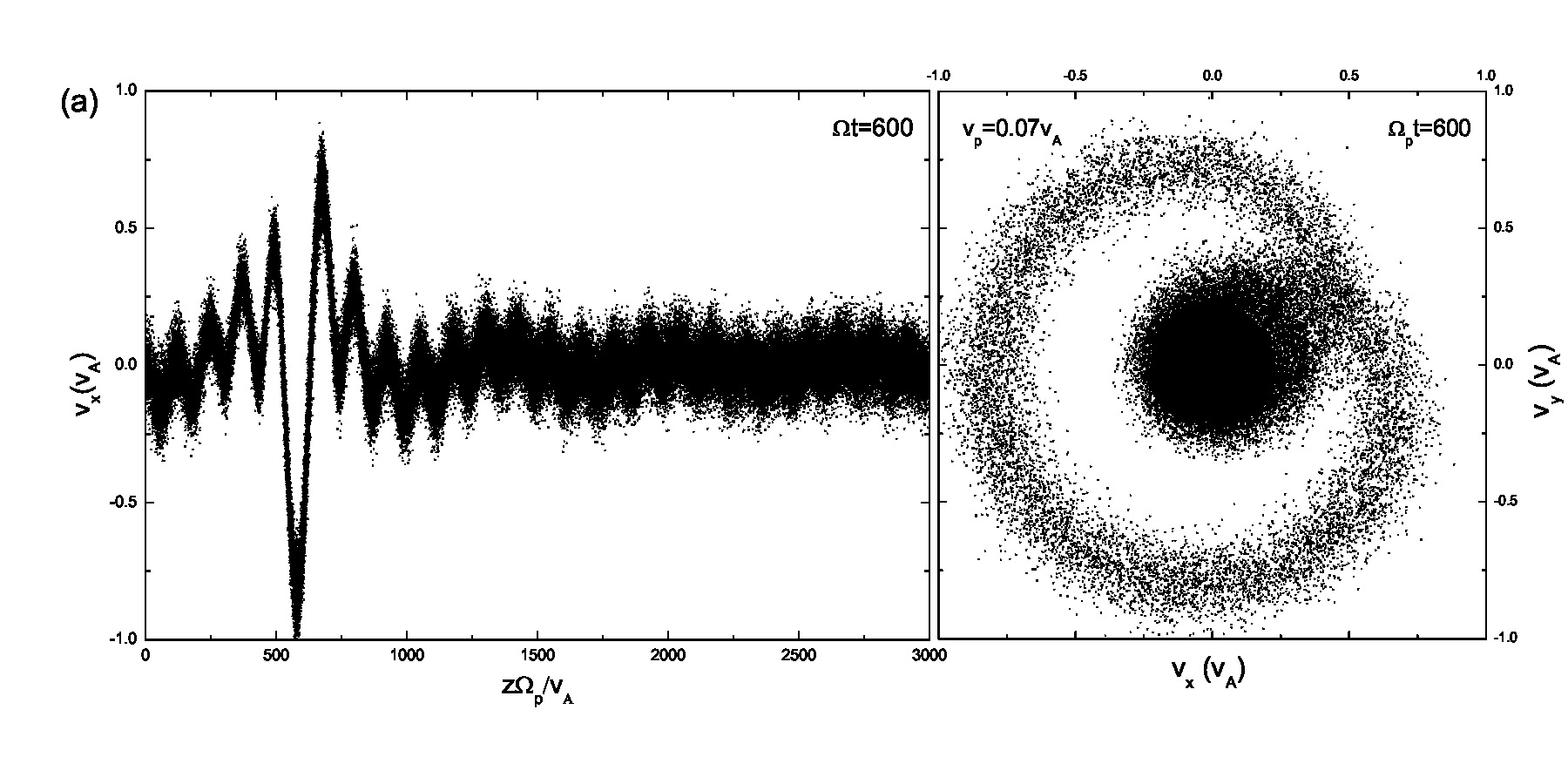} 
\caption{Proton scatter plots of a spectrum of circularly polarized Alfv\'{e}n waves with
same initial phases $\protect\varphi_k$ at $\Omega _{p}$t=600; $v_{p}$=0.07$v_{A}$.}
\label{fig8}
\end{figure}

\begin{figure}[tbp]
\centering
\includegraphics[scale=0.35]{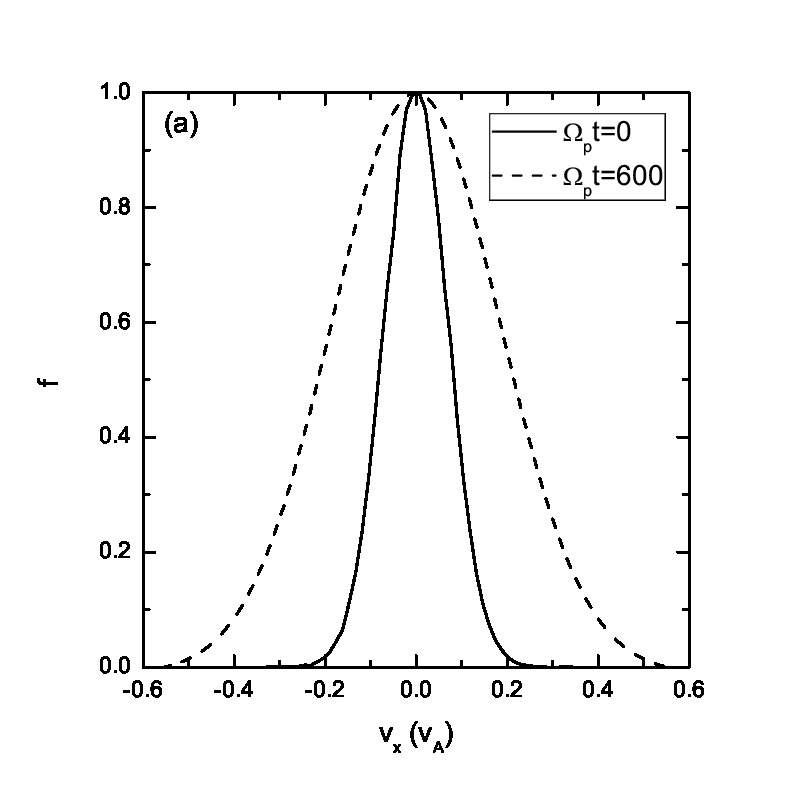} 
\includegraphics[scale=0.35]{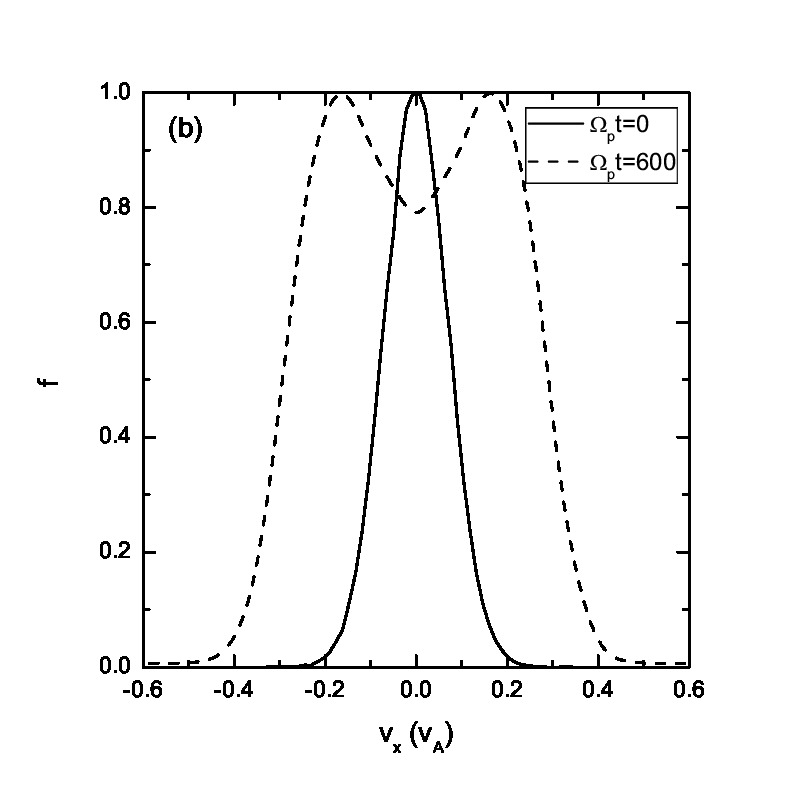}
\includegraphics[scale=0.35]{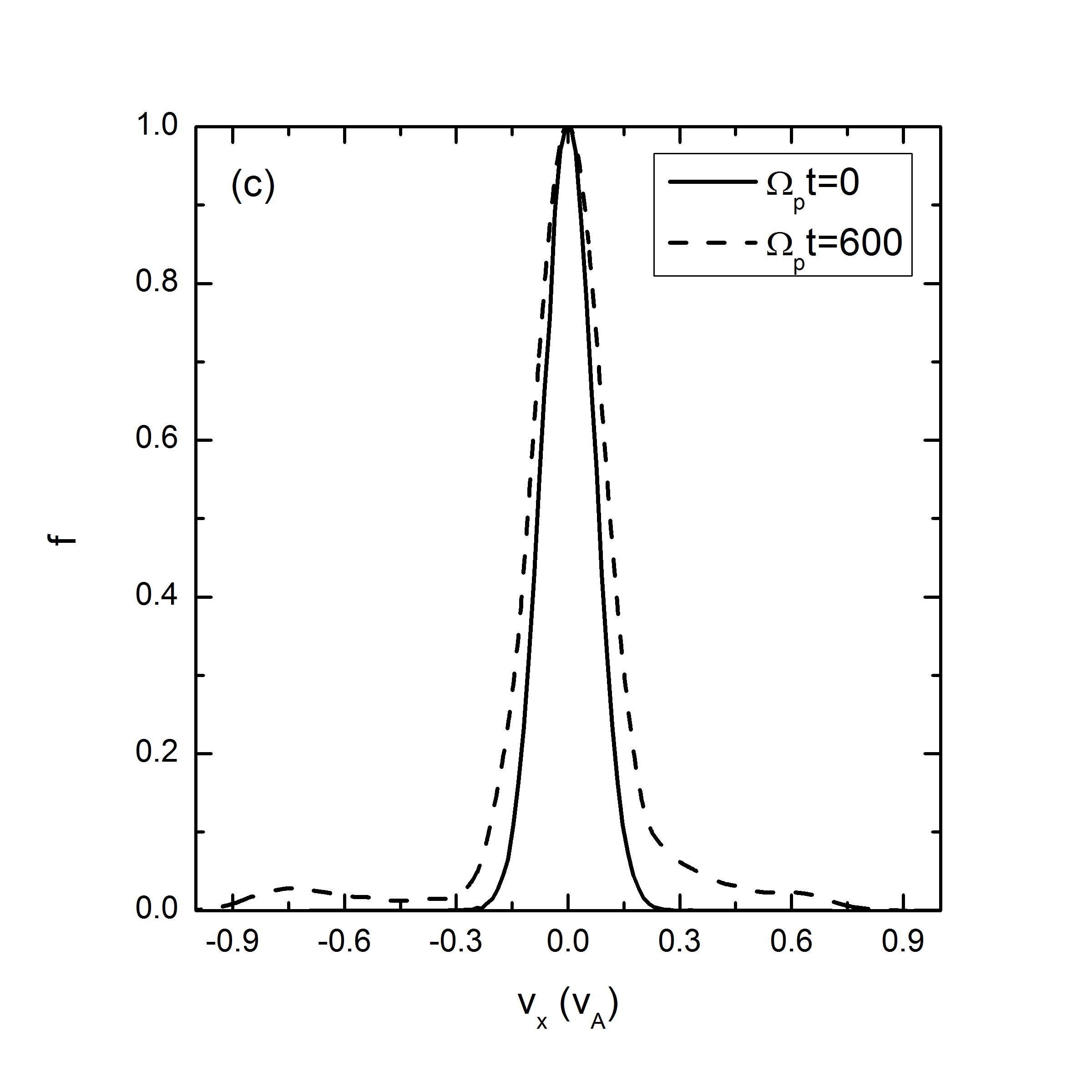}
\includegraphics[scale=0.35]{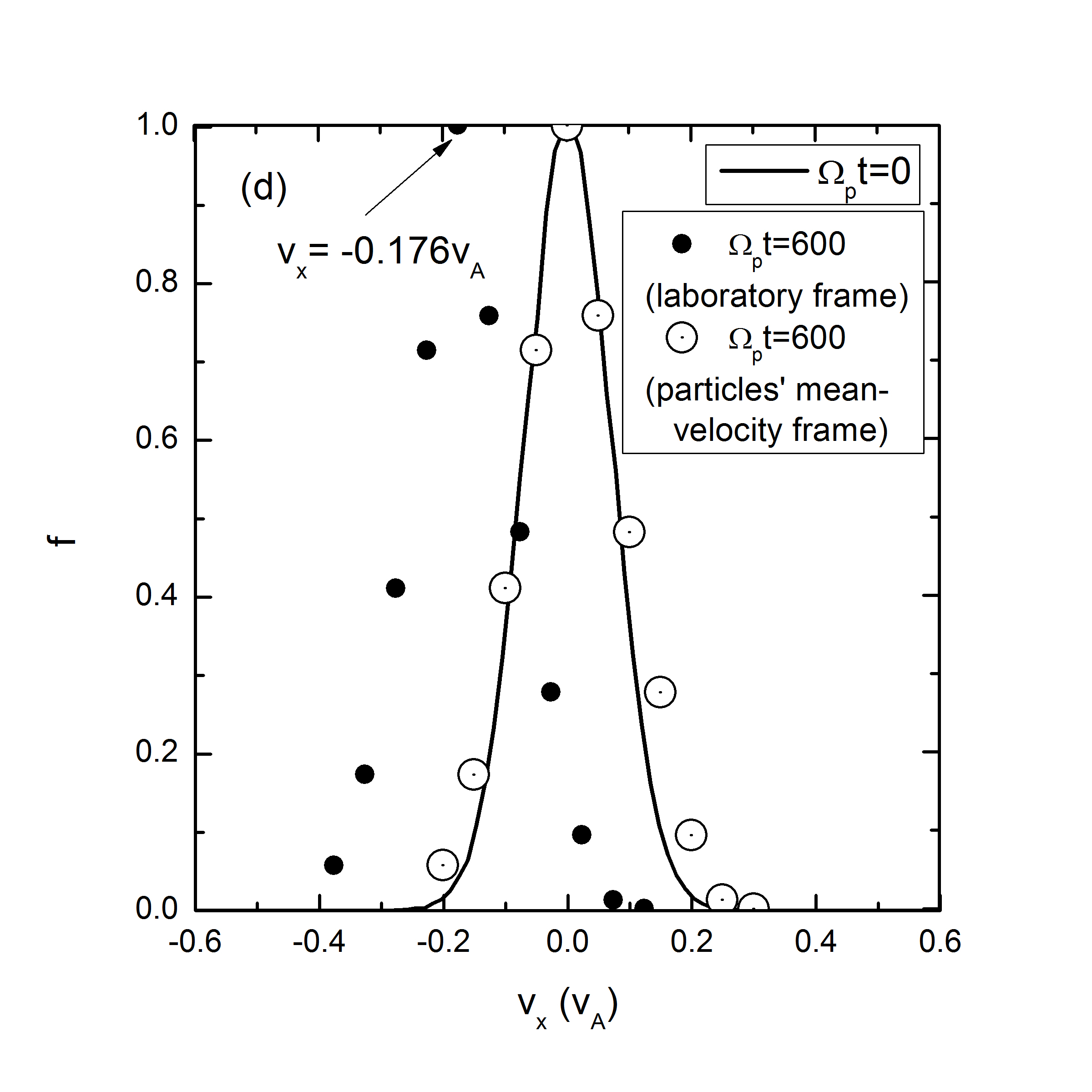}
\caption{The normalized velocity distribution functions plotted against $v_x$
at $\Omega _{p}$t=0 (solid line) and 600 (dashed line). (a) A spectrum of
circularly polarized Alfv\'{e}n waves with random phases $\protect\varphi_k$%
; (b) A circularly polarized monochromatic Alfv\'{e}n wave. $v_{p}$=0.07$%
v_{A}$; (c) A spectrum of circularly polarized Alfv\'{e}n waves with same initial
 phases $\protect\varphi_k$. (d) The local normalized velocity distribution function
based on the statistics of protons in the spatial range $1490<z\Omega_p/v_A<1510$
under a spectrum of circularly polarized Alfv\'{e}n waves with random phases 
$\protect\varphi_k$. The black dots denote the shifted local normalized velocity 
distribution function in the laboratory frame while the white dots represent the 
local normalized velocity distribution function in the particles' mean-velocity frame. The local 
statistical mean velocity equals to $v_x(1490<z\Omega_p/v_A<1510)=-0.176v_A$.  }
\label{fig4}
\end{figure}

Compared with the ion behavior under a monochromatic wave, the ion motion
tends to become random under a wave spectrum as indicated in Figs.\ref{fig21} 
\& \ref{fig2}. Fig.\ref{fig21} presents the velocity scatter plots 
of test particles in the $v_{x}-v_{y}$ space for a spectrum of circularly 
polarized Alfv\'{e}n waves with random phases $\varphi _{k}$. When the thermal
speed is two orders of magnitudes smaller than the Alfv\'{e}n speed $v_{A}$ (i.e., $%
v_{p}$=0.01$v_{A}$ and 0.03$v_{A}$), test particles cannot fully fill the
circle in the $v_{x}-v_{y}$ velocity space. However, with the increase of
the initial thermal speed $v_{p}$, the circle in the velocity space is
fully filled with test particles with different velocities. Figs.\ref{fig21}(e)\&(f) 
show the velocity scatter plots when adopting 1001 wave modes. Compared with 
Figs.\ref{fig21}(a)\&(b), Figs.\ref{fig21}(e)\&(f) show that protons tend to
fully fill the circle in the velocity space, indicating that the 
ion velocity distribution tends to be a full-filled circle in the velocity
space when the number of wave modes, $N$, is large enough, regardless of the relatively
small thermal speed. It can be observed based on the comparison between Fig.\ref{fig31} 
and Fig.\ref{fig21} as well. The phase space proton scatter diagrams by adopting a wave 
spectrum with wave modes $N$=41 are shown in Fig.\ref{fig2} . The results, however, 
reveal significantly different particle distributions in the phase space compared 
with those shown in Fig.\ref{fig3}. The main conclusion drawn from Fig.\ref{fig2} 
is that the velocity fluctuations caused by wave activity is quasirandom, and 
thereby could mimic the real heating (also refer to Fig.\ref{fig4}). However, as indicated 
in both Fig.\ref{fig11} and Fig.\ref{fig2}, the pseudoheating caused by these wave activities is reversible, 
indicating no dissipation of wave fields, and therefore does not represent real heating 
in thermodynamic sense. It is also interesting to investigate the ion behavior
under a spectrum of circularly polarized Alfv\'{e}n waves with same initial phases $\protect\varphi_k$. 
It is noteworthy that the wavelength and wave frequency among different wave modes are still different. 
Fig.\ref{fig8} illustrates the scatter plots of the test particles in the $v_x-z$ and $v_x-v_y$ space.
There is a pulse-like structure in the protons' phase space distribution due to the coherence of wave modes. The
proton distribution in the $v_x-v_y$ space is primarily consist of two parts: the broadening core Maxwellian
distribution and the outer ring structure. The broadening is caused by the wave forces (or their spectra)
while the accelerated particles in the outer ring result from the Alfv\'{e}nic turbulence with phase coherent 
wave modes as indicated in Ref.\cite{ra1}, where the acceleration of charged particles by large amplitude 
MHD waves was studied. In contrast, if Alfv\'{e}nic turbulence without any envelope modulation\cite{ra2} is given,
the ``acceleration'' may not be observed. These high energy particles in the outer ring may escape from the region 
of interaction with the Alfv\'{e}n waves and can contribute to the fast particle population in astrophysical and space plasmas.

In Fig.\ref{fig4}, we present the normalized velocity distribution function
at different time $\Omega _{p}$t=0 (solid line) and 600 (dashed line). 
The broadening of the distribution functions shown in Figs.\ref{fig4}(a)-(c) 
are based on the statistics of all the particles with different spatial coordinates 
$z$ while Fig.\ref{fig4}(d) is based on the local statistics of protons in the spatial range 
$1490<z\Omega_p/v_A<1510$, thus being a local normalized velocity
distribution function around $z\Omega_p/v_A=1500$ as indicated by
the line in Fig.\ref{fig2}(d). The velocity spreading in Figs.\ref{fig4}(a)-(c) is caused 
by averaging over the wave effects. 
Inspection of Figs.\ref{fig4}(a)-(b) reveals that the Maxwellian distribution is
more likely to maintain under a wave spectrum compared with a 
monochromatic Alfv\'{e}n wave due to the fact that the ion motion under
a wave spectrum is quasirandom, which is in good agreement with the analytic 
derivation by Wu and Yoon using the quasi-linear theory\cite{wuPRL}. 
It also points towards different kinetic behaviors of particles under a wave 
spectrum and a monochromatic Alfv\'{e}n wave. It is well known that 
enhanced Alfv\'{e}nic turbulences exist pervasively in the solar wind and interplanetary 
space, while in astrophysical observation measurements of temperature are based on 
spectroscopic data collected from the source region of interest. Therefore it is very difficult 
to distinguish the real heating from the pseudoheating due to the restriction of 
spatial resolution of the instruments and the presence of wave 
forces (or their spectra). The suprathermal tail and small bump shown 
in Fig.\ref{fig4}(c), where the $v_x$ axis scale is different from Figs.\ref{fig4}(a)-(b),
corresponds to the accelerated (suprathermal) ions 
due to the wave modes coherence. The bump-on-tail structure may excite
plasma instabilities, however, the detailed discussion of these instabilities is beyond the scope
of this paper. It also needs to point out that here we only consider the pseudoheating
and investigate the normalized velocity distribution function corresponding to this process
while the real heating via nonresonant interaction with Alfv\'{e}n waves always
coexists\cite{wangPRL,Dongr}. The velocity distribution function, therefore, is supposed to be
broader than that shown in Fig.\ref{fig4} and it is possible that the broadening 
effects smooth the velocity distribution function and eliminate the bump on the tail.
In Fig.\ref{fig4}(d), we investigate the local normalized velocity 
distribution function based on the statistics of protons in the spatial range $1490<z\Omega_p/v_A<1510$.
If the spatial range is too small, there will be insufficient particles to be counted for statistics.
The local statistical mean velocity equals to $v_x$=-0.176$v_A$ which agrees well with the 
result shown in Fig.\ref{fig2}(d). The local normalized velocity distribution function in
the particles' mean-velocity frame at $\Omega_pt=600$ (as indicated by the white dots) is in good agreement with the initial Maxwellian
distribution based on Eq.(\ref{MaxDis}). The slight difference between these two distribution functions results from
the fact that the local statistics are based on a small spatial range ($1490<z\Omega_p/v_A<1510$) and therefore it
can be potentially affected by the effects of wave spectra. This leads to a slightly broader local velocity distribution. 
The result shown in Fig.\ref{fig4}(d) indicates that the analytic theory in Sec.2 is on the basis of the local equilibrium velocity distribution of the ions.

\vskip 10mm


\begin{figure}[tbp]
\centering
\includegraphics[scale=0.3]{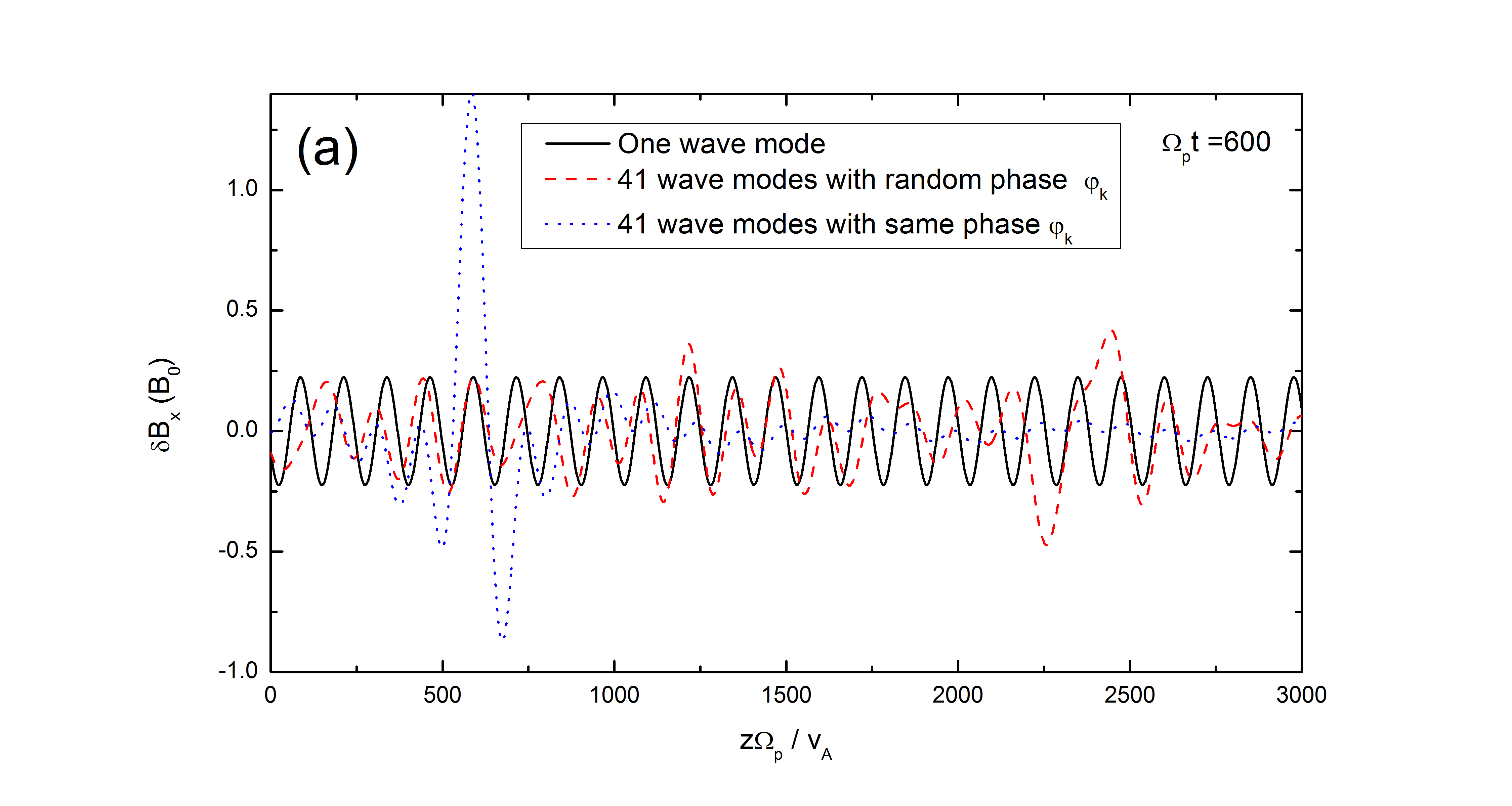} %
\includegraphics[scale=0.3]{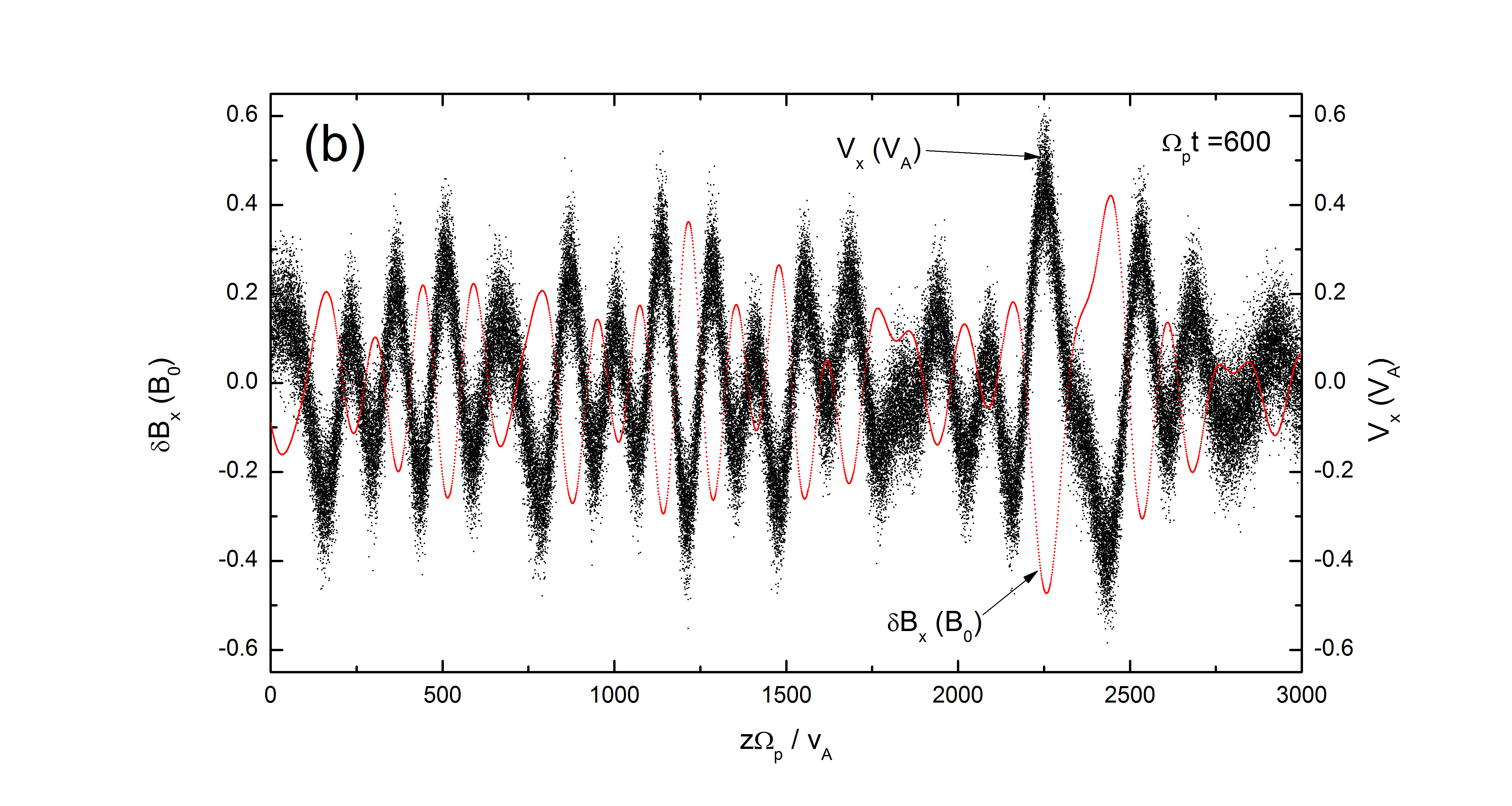}
\includegraphics[scale=0.3]{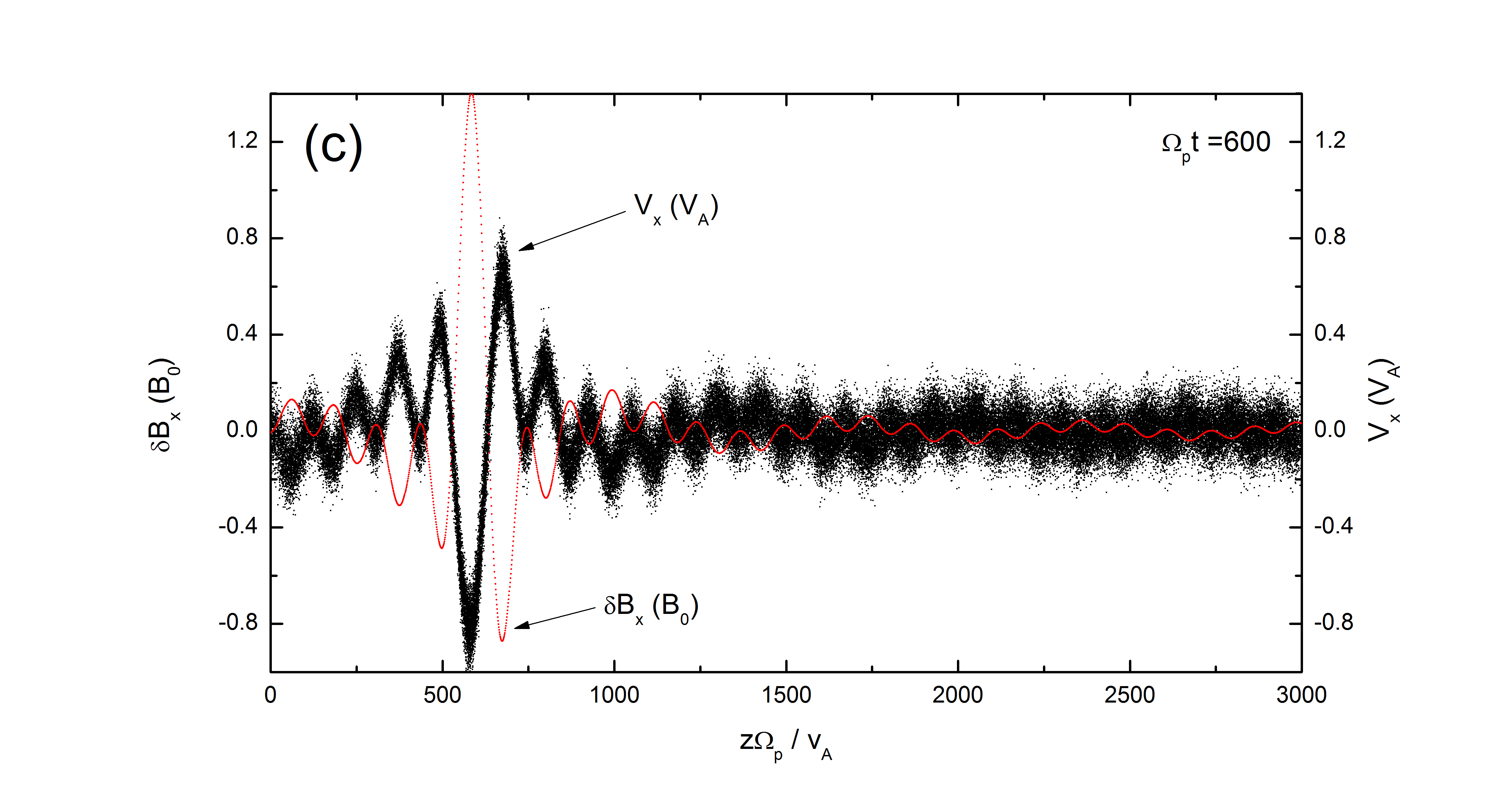}
\caption{ (Color online) (a) The spatial distribution of Alfv\'{e}n wave magnetic fields
at $\Omega _{p}$t=600. (b)-(c) The Alfv\'{e}n wave magnetic field spatial distribution and 
the corresponding scatter plot in the $v_x-z$ phase space at $\Omega _{p}$t=600;
(b) the case with random phases $\protect\varphi_k$ and (c) the case with same phases $\protect\varphi_k$. }
\label{fig42}
\end{figure}

The pseudoheating becomes important when the magnetic perturbation ($\delta B$) 
is relatively large compared with the background magnetic field ($B_0$).
Fig.\ref{fig42} illustrates the spatial distribution of the Alfv\'{e}n wave magnetic
fields with different wave modes, $N$, and the corresponding scatter plot in the
$v_x-z$ phase space. The solid curve for the magnetic field of a monochromatic
Alfv\'{e}n wave is simply a sinusoidal wave. The dashed and dotted curves represent a
spectrum of Alfv\'{e}n waves with 41 wave modes that has random and same phase $%
\varphi _{k}$, respectively. They can mimic the Alfv\'{e}nic turbulence due to the quasirandom 
fluctuations of the wave fields. The amplitude of the
magnetic perturbation can be quite large locally, thus may produce some energetic 
particles that can escape from the constraint of the Alfv\'{e}n waves 
and contribute to fast particle population in astrophysical and space 
plasmas. As indicated in Fig.\ref{fig42}, the larger the magnetic
perturbations are, the more effective the distribution functions are broadened.  
Refs.\cite{LS,AA} suggest that small scale reconnection events occur
during the solar flares, which can provide large magnitude spike-like 
magnetic field fluctuations. Additionally, the cluster observation of surface 
waves in the ion jets from magnetotail reconnection also shows that $\delta B/|B|$ 
can be as high as 0.5 and occasionally even higher\cite{dai2}. Furthermore, the
amplitude of the magnetic perturbation associated with Alfv\'{e}n waves can be even 
larger than the ambient magnetic field in certain cases observed by the 
\emph{Wind} satellite\cite{wind}. Figs.\ref{fig42}(b)\&(c) also show that the distribution 
of test particles in the $v_{x}$-$z$ phase space and the wave field perturbations are
in antiphase ($\pi$ phase difference) due to the fact that the 
proton motion is parasitic to Alfv\'{e}n waves, indicating an exchange of
energy between the particles' kinetic energy and the magnetic energy. 
Therefore Fig.\ref{fig42}(b) indicates that the psuedoheating is a consequence of 
equilibrium MHD system.

It is noteworthy that the parallel electric field and wave damping are not included in the test particle simulations of our current work. In the self-consistent simulations such as hybrid simulations (i.e., kinetic description of ions, fluid electrons), the phase-correlation among the Fourier modes could affect, for instance, that the ponderomotive force resulting from the envelope-modulated Alfv\'{e}n waves heat ions through the nonlinear Landau damping\cite{r3} and the parallel heating of ions due to the nonlinear Landau damping\cite{r4}.  In Ref.\cite{r3}, the ponderomotive force and beat interaction are identified as the most important nonlinear effects in proton heating by nonlinear field-aligned Alfv\'{e}n waves in the solar coronal holes. Interestingly, they found that the nonlinearity is particularly strong when the wave spectrum consists of counterpropagating modes of equal intensity, even if the intensity is relatively low. Moreover, from the hybrid approach, dissipation processes of the Alfv\'{e}nic turbulence with the broadband wave number spectrum can be different from those of monochromatic Alfv\'{e}n waves, since the former is associated with the density fluctuations, $|\delta n/n|$, and the resultant spatial modulation of $|B|^2$ due to compressive effects of ponderomotive forces\cite{r5,r6}: right-hand polarized Alfv\'{e}nic turbulence with such an envelope modulation can be dissipated due to the nonlinear Landau damping, while left-hand polarized Alfv\'{e}nic turbulence with the broadband spectrum is preferentially dissipated by the modulational instability. On the other hand, low-frequency, monochromatic Alfv\'{e}n waves are relatively stable to the linear collisionless dissipations such as Landau damping and cyclotron damping, nonlinear wave-wave interactions parametric instabilities are important for the dissipation of these waves. The decay instability is dominant for dissipation of the right-hand polarized finite-amplitude monochromatic Alfv\'{e}n waves in low $\beta$ plasmas, while the left-hand polarized waves can also be dissipated via the modulational instability\cite{r4}. These wave damping mechanisms are closely related to the present work; therefore, in order to study the characteristics of the ``acceleration'' and ``heating'' process in more detail, comprehensive and self-consistent studies are necessary in the future.

\section{CONCLUSION}

Ion pseudoheating by low-frequency Alfv\'{e}n waves is investigated based on the
comparison between a monochromatic Alfv\'{e}n wave and a wave spectrum. 
Both analytic and simulation results show that $E\times B$ drift plays 
a principal role in this process and the proton motion
is parasitic to Alfv\'{e}n waves. It indicates the psuedoheating is a consequence of
the equilibrium in the MHD system. Our results are in good agreement
with the previous studies based on pitch-angle scattering. More importantly, 
it provides a simple understanding of the reversible property of this process from $E\times B$ drift 
point of view; if wave magnetic and electric fields disappear, there will 
be no drift velocity $v_{E}$ and therefore no pseudoheating. We showed 
that the wave spectra contribute to the broadening of the Maxwellian 
distribution during the pseudoheating, and it is therefore difficult to 
exclude the apparent temperature $T_p$ from observations due to the low 
spatial resolution of the instruments. It is of particular interests to note that the Maxwellian 
shape is more likely to maintain during the pseudoheating under a wave spectrum 
compared with a monochromatic Alfv\'{e}n wave. We, therefore can conclude that
the kinetic behavior of ions under a monochromatic wave and a wave spectrum 
is totally different. Moreover, we illustrated that $E \times B$ drift can produce 
energetic particles under a spectrum  of Alfv\'{e}n waves, which may 
contribute to fast particle population in astrophysical and space plasmas.

\vskip 10mm

\noindent{\large \textbf{Acknowledgments:}} C.F. Dong appreciates many
fruitful discussions with Prof. C.B. Wang and Prof. Y.Y. Lau. The authors 
would like to thank the anonymous referees' helpful comments and suggestions.

\end{document}